\documentclass{jfm}
\usepackage{epstopdf}
\usepackage{natbib}
\usepackage{amsmath}
\usepackage{enumitem}
\usepackage{bm}
\usepackage{color}
\usepackage[parfill]{parskip}
\makeatletter
\g@addto@macro\normalsize{%
  \setlength\abovedisplayskip{6pt}
  \setlength\belowdisplayskip{6pt}
  \setlength\abovedisplayshortskip{6pt}
  \setlength\belowdisplayshortskip{6pt}
}
\makeatother
\newcommand\Ca{\mbox{\textit{Ca}}}
\newcommand\Mrat{\mbox{\textit{M}}}
\newcommand\Rrat{\mbox{\textit{R}}}
\newcommand{\iu}{{\mbox{\text{i}}}}
\newcommand{\expf}{{\mbox{\text{exp}}}}
\allowdisplaybreaks

\begin{document}

\newtheorem{lemma}{Lemma}
\newtheorem{corollary}{Corollary}

\shorttitle{Interfacial instability of lubricating liquid films sheared by a gas flow} 
\shortauthor{Mikl\'os V\'ecsei, Mathias Dietzel and Steffen Hardt} 

\title{Interfacial instability of liquid films coating the walls of a parallel-plate channel and sheared by a gas flow}

\author
 {
 Mikl\'os V\'ecsei\aff{}
  \corresp{\email{vecsei@csi.tu-darmstadt.de}},
  Mathias Dietzel\aff{}
  \and 
  Steffen Hardt\aff{}
  }
\affiliation
{
\aff{}
Institute for Nano- and Microfluidics, Center of Smart Interfaces, Technische Universit\"at Darmstadt, Alarich-Weiss-Stra{\ss}e 10, 64287 Darmstadt, Germany
}

\date{2015.12.08}

\maketitle

\begin{abstract}
The stability of liquid films coating the walls of a parallel-plate channel and sheared by a pressure-driven gas flow along the channel centre is studied. The films are susceptible to a long-wavelength instability, whose dynamic behaviour is found - for sufficiently low Reynolds numbers and thick gas layers - to be described by two coupled non-linear partial differential equations. To the best of our knowledge, such coupled fully non-linear equations for the film thicknesses have not been derived previously. A linear stability analysis conducted under the condition that the material properties and the initial undisturbed liquid film thicknesses are equal can be utilized to determine whether the interfaces are predominantly destabilized by the variations of the shear stress or by the pressure gradient acting upon them. The analysis of the weakly non-linear equations performed for this case shows that instabilities corresponding to a vanishing Reynolds number are absent from the system. Moreover, for this configuration, the patterns emerging along the two interfaces are found to be identical in the long-time limit, implying that the films are fully synchronized. A different setup, where the liquid films have identical material properties but their undisturbed thicknesses differ, is studied numerically. The results show that even for this configuration the interfacial waves remain phase-synchronized and closely correlated for a broad range of parameters. These findings are particularly relevant for multiphase flow in narrow ducts, for example in the respiratory system or in microfluidic channels.
\end{abstract}

\section{Introduction}

The emergence of flow instabilities in stratified systems is common in nature. One of the most well-known related phenomenon is the Kelvin-Helmholtz instability, appearing on the initially flat interface between two inviscid fluid layers flowing parallel to each other. If the velocity of the two fluids differ, the system may be unstable to small disturbances in the flow field \citep{drazin_hydrodynamic}, which are escalated by the corresponding variations in the dynamic (Bernoulli) pressure and grow into a vortex sheet \citep{chandrasekhar_hydrodynamic}. If the viscous stresses are not negligibly small, they may also trigger the formation of patterns at interfaces, for which the formation of ripples on sand beds sheared by a liquid flow \citep{charru_sand_ripple} is an example. Therefore, for general co-current fluid flows, it is reasonable to consider the variations in both, the pressure and the viscous stresses, in order to obtain the time evolution of the system. Since the early discussion of \cite{yih_viscous_strati}, the instability of the parallel flow of fluids with different viscosities has been the subject of numerous papers, illustrating the considerable challenge their analysis often poses. 

This paper will focus on the instability induced by a planar gas flow between two thin liquid films coating flat walls opposite to each other. Such systems are important for a number of applications. For instance, corresponding flow patterns appear during the boiling of a liquid in microchannels \citep{kandlikar_flow_boiling}, and the understanding of interfacial instabilities in gas-liquid flows is also vital for the mapping of the different flow regimes in confined multiphase systems. These maps have been explored analytically \citep{taitel_anal_two_phase_map}, experimentally (\cite{saisorn_exp_two_phase_map}, \cite{triplett_anal_two_phase_map}) and numerically \citep{talimi_numeric_two-phase}. Furthermore, \cite{heil_airway_closure} and \cite{johnson_airway_collapse} have used core-annular gas flow as a model system for the discussion of the closure of airways. Although core-annular flows have in general different properties than planar flows, their features are similar as long as the tube curvature and the corresponding capillary effects are negligibly small \citep{renardy_triple_layer}. 

For thin liquid films, interfacial instabilities are strongly influenced by surface tension. The stabilizing effect of capillarity dampens sharp deformations of the interface, so that for sufficiently thin films the evolution of the instabilities can be described with the long-wavelength approximation. The analyses of \cite{hooper_double_nonlin} and \cite{shlang_double_nonlin} have shown that the co-current flow of two superposed liquid films with large interfacial tension is governed by the Kuramoto-Sivashinsky equation. In contrast to the relative compactness of this non-linear partial differential equation, its analysis is remarkably challenging. Interfacial instabilities occurring in the planar flow of three superposed liquid films have been first discussed by \cite{li_three_layer}. It was found that in contrast to double-layer configurations, these systems can become unstable even if inertial effects are negligible. Using linear stability analysis, this phenomenon has also been studied in detail by \cite{kliakhandler_alpha}. However, the large influence of the non-linear terms arising from advective effects severely limits the range of validity of the linearized problem. \cite{papaef_2013} argued that the weakly non-linear equations of stratified films should take the form of coupled Kuramoto-Sivashinsky equations. Nonetheless, an advective term appearing in the equations complicates the analytical description of the interfaces. These novel instabilities in multilayer systems, now commonly referred to as kinetic instabilities, were found to be either the result of a resonance-like coupling between the interfaces or of a forth-order generalization of the Majda-Pego instability \citep{majda-pego_instability}.

The studies of \cite{kliakhandler_alpha} and \cite{papaef_2013} focused on systems utilizing fluids of arbitrary viscosities. The mathematical complexity of such systems hinders their analytical description, limiting the investigations to either numerical simulations or to the discussion of very simple configurations. However, for the long-wavelength instability due to planar gas flow between thin liquid films, being of extraordinary relevance for a number of practical applications, one may - based on the large discrepancy between the viscosities of the liquid and the gas phase - simplify the governing equations. To the best of our knowledge, such an analysis has not yet been performed for these three-layer systems and will be the focus of this paper. Section \ref{sec:ev_eq} discusses the derivation of the evolution equations for the film thicknesses. Section \ref{sec:sym_sys} focuses on symmetrical systems, where the compact form of the equations and the small number of independent parameters permits a deeper analytical treatment. Finally, in section \ref{sec:asym_sys}, a short qualitative summary is given of the results obtained by numerically simulating asymmetrical systems.

\section{Evolution equations of the interfaces}\label{sec:ev_eq}
\subsection{Governing equations}\label{sec:gov_eq}
\begin{figure}
\centerline{\includegraphics[width=0.55\linewidth]{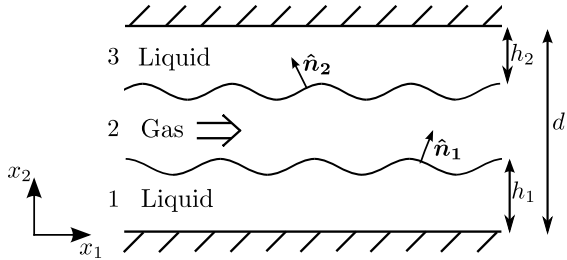}}\caption{Illustration of gas flow between two liquid films.}\label{fig:triple_layer}
\end{figure}
This section focuses on the derivation of the evolution equations for the gas and liquid film flows. The analysis in two spatial dimensions is restricted to immiscible, incompressible, Newtonian fluids, i.e. the latter two assumptions imply sufficiently small flow velocities and shear rates, respectively. Furthermore, the flow is assumed to be driven solely by an externally imposed pressure gradient. A more detailed derivation of the general set of the governing equations is given by \cite{papaef_2013}. According to the notations introduced in figure \ref{fig:triple_layer}, the momentum equations in the bulk fluids and at the interfaces are given by
\begin{align}
\rho_i \frac{d\bm{v}}{dt}&=-\bm{\nabla} \cdot \mathsfbi{P}_{\bm i},\label{eq:bulk_mom}\\
0&= - \gamma_i\left(\bm{\nabla_{i}} \cdot \bm{\hat{n}_i}\right)\bm{\hat{n}_i}+\left(\mathsfbi{P}_{\bm i}-\mathsfbi{P}_{\bm i+1}\right)\bm{\hat{n}_i},\label{eq:surf_mom}
\end{align}

\noindent where the liquid films are assumed to be sufficiently thin to neglect the gravitational force. The subscript $i$ defines the fluid film or the interface the physical quantity is referring to. In this sense the three phases are numbered "1", "2" and "3" from bottom to top, and in a similar manner, the lower interface is denoted with a "$1$" and the upper one with a "$2$" (see figure \ref{fig:triple_layer}). The gradient along interface $i$ is denoted by $\bm{\nabla_i}$, and its normal vector by $\bm{\hat{n}_i}$. The directions of $\bm{\hat{n}_i}$ are defined in figure \ref{fig:triple_layer}. The symbols $\rho_i$ and $\gamma_i$ refer to the mass densities of the fluids and the constant surface tensions, respectively. The velocity field is denoted by $\bm{v}=(v_1,v_2)$, and its substantial time derivative by $d\bm{v}/dt$. The stress tensors, $\mathsfbi{P}_{\bm i}$, are composed of the scalar hydrodynamic pressures $p_i$ and the viscous stress tensors, namely

\begin{equation}
\begin{aligned} \mathsfbi{P}_{\bm i}&=p_{i}\mathsfbi{I}+\mathsfbi{E}_{\bm i}, &\\
\left[\mathsfbi{E}_{\bm i}\right]_{kl}&=-\mu_i\left(\frac{\partial v_k}{\partial x_l}+\frac{\partial v_l}{\partial x_k}\right) \quad\quad \text{ }k,l\in[1,2],
\end{aligned}
\end{equation}

\noindent where the identity matrix is referred to by $\mathsfbi{I}$, while $\mu_i$ denote the dynamic viscosities.

For the non-dimensionalisation of the equations one introduces $\lambda_{\text{char}}$ as the - still unknown - characteristic wavelength of the variation of the flow field along the horizontal direction. Consequently, the $x_1$ coordinate is non-dimensionalised by $X=x_1/\lambda_{\text{char}}$. The quantities $Y=x_2/d$ and $H_i=h_i/d$ are introduced as the dimensionless vertical coordinate and film thickness, respectively, while the separation distance between the walls is given by $d$. The volume flux of the gas flow at the entrance of the channel, $Q$, is utilized for defining the dimensionless velocities $\bm{V}=(V_X,V_Y)=(v_1,v_2/\epsilon)d/Q$. In this definition the $\epsilon=d/\lambda_{\text{char}}$ multiplier of the vertical velocity originates from the different spatial scalings applied in the horizontal and the vertical direction. The rescaled time variable reads $\tau=tQ/(\lambda_{\text{char}}d)$. The dimensionless pressures are given by $P_i=p_i\epsilon d^2/(\mu_2 Q)$, with $\mu_2$ being the dynamic viscosity of the gaseous medium. Lastly, the viscosity ratios $\Mrat_i=\mu_i/\mu_2$ and density ratios $\Rrat_i=\rho_i/\rho_2$ are introduced. With these new variables the momentum equations take the form

\begin{equation}
\begin{aligned}
\epsilon \Rey \Rrat_i\frac{dV_X}{d\tau}=&\Mrat_i\left(\epsilon^2\frac{\partial^2 V_X}{\partial X^2}+\frac{\partial^2 V_X}{\partial Y^2}\right)-\frac{\partial P_i}{\partial X},\\
\epsilon^3 \Rey \Rrat_i\frac{dV_Y}{d\tau}=&\Mrat_i\left(\epsilon^4\frac{\partial^2 V_Y}{\partial X^2}+\epsilon^2\frac{\partial^2 V_Y}{\partial Y^2}\right)-\frac{\partial P_i}{\partial Y}.
\end{aligned}\label{eq:nondim_bulk_ns}
\end{equation}

\noindent The Reynolds number is denoted by $\Rey=\rho_2Q/\mu_2$. The interfacial momentum equations in the tangential and the normal directions are given by

\begin{equation}
\begin{aligned}
\Mrat_{i+1}\left(\epsilon^2\frac{\partial V_Y}{\partial X}{\Bigg|}_{i+1}+\frac{\partial V_X}{\partial Y}{\Bigg|}_{i+1}\right)&-\Mrat_i\left(\epsilon^2\frac{\partial V_Y}{\partial X}{\Bigg|}_i+\frac{\partial V_X}{\partial Y}{\Bigg|}_i\right)=\\
&-\frac{4 \epsilon^2\partial H_i/\partial X}{1-\epsilon^2 (\partial H_i/\partial X)^2}\left(\Mrat_{i+1}\frac{\partial V_Y}{\partial Y}{\Bigg|}_{i+1}-\Mrat_i\frac{\partial V_Y}{\partial Y}{\Bigg|}_i\right),\\
P_{i+1}-P_i-2\epsilon^2\frac{1+\epsilon^2(\partial H_i/\partial X)^2}{1-\epsilon^2(\partial H_i/\partial X)^2}&\left(\Mrat_{i+1}\frac{\partial V_Y}{\partial Y}{\Bigg|}_{i+1}-\Mrat_i\frac{\partial V_Y}{\partial Y}{\Bigg|}_i\right)=\\
&\frac{\epsilon}{\Ca_i}\frac{\partial^2H_i/\partial X^2}{[\epsilon^2(\partial H_i/\partial X)^2+1]^{3/2}}.
\end{aligned}
\end{equation}

\noindent The Capillary numbers are denoted by $\Ca_i=\mu_2Q/(\epsilon^2d\gamma_i)$. According to the arguments of \cite{ooshida_1999} for the related problem of falling-film flows, the $\epsilon^2$-term in the definition of the Capillary number ensures that the scaling is consistent, namely that the characteristic length scale of the non-dimensional patterns are only dependent on $\Ca$ and $\Rey$, and independent of $\epsilon$. This is apparent from expression \eqref{eq:nondim_wavenum}.

Finally, the dimensionless continuity equation in the layers read
\begin{equation}
\frac{\partial V_X}{\partial X}+\frac{\partial V_Y}{\partial Y}=0,
\end{equation}
\noindent while the kinematic condition along the interfaces leads to
\begin{equation}
\begin{aligned}
\frac{\partial H_1}{\partial \tau}+V_X{\Big|}_{H_1}\frac{\partial H_1}{\partial X}-V_Y{\Big|}_{H_1}&=0,\\
\frac{\partial H_2}{\partial \tau}+V_X{\Big|}_{1-H_2}\frac{\partial H_2}{\partial X}+V_Y{\Big|}_{1-H_2}&=0.\\
\end{aligned}\label{eq:cont_int}
\end{equation}

\subsection{Framework of the analysis}\label{sec:framework}

The mathematical analysis of the full equations poses a considerable challenge. Therefore, it is advantageous to introduce a few simplifications.

\begin{enumerate}
\setlength{\itemindent}{2.78cm}
\item[\textbf{Assumption 1}]\hspace{4pt} The paper focuses on interfacial instabilities of thin liquid films. The dominant influence of capillarity and viscosity on the behaviour of  such systems dampens short-wavelength deformations and instabilities. Therefore, the long-wavelength approximation \citep{oron_lubrication} is utilized for the analysis. Mathematically this corresponds to writing all quantities in terms of a power series in $\epsilon$ and neglecting the second or higher order terms in this parameter.
\item[\textbf{Assumption 2}]\hspace{4pt} For inertialess systems ($\Rey \rightarrow 0$) the simplified equations are reducible to two equations describing the movement of the interfaces. In general, the $\Rey\ll 0$ condition is not necessarily fulfilled, so that the inertial term in the momentum equation becomes important. To circumvent this problem, the analysis will be restricted to configurations where $O(\epsilon^2\Rey^2)\ll1$. Formally, one assumes that a first-order perturbation in $\epsilon\Rey$, linearizing the governing equations, gives a sufficiently accurate description of the flow field. This approach has been successfully applied many times, for example to the related problem of liquid film stability during spin coating \citep{davis_spin_coat}.
\item[\textbf{Assumption 3}]\hspace{ 4pt} In order to reduce the set of equations to only two evolution equations, it is sufficient to apply the previous two assumptions. Nevertheless, conducting the necessary calculations is cumbersome and is often only possible numerically. Thus, one may lose some of the physical insight due to the mathematical complexity of the problem. However, for the stratified liquid-gas-liquid flow one may take advantage of the circumstance that the gaseous layer has a considerably smaller viscosity than the liquid, so that $(\Mrat_1)^2, (\Mrat_3)^2\gg1$. Hence, the assumption is made that all terms containing second or higher orders of $ 1/M_1$ and $1/M_3$ can be neglected from the equations. As reasoned below, in this limit the films can be considered as being semi-rigid.
\end{enumerate}

The mathematical implementation of these assumptions is straightforward. Assuming that all dependent variables can be expanded into Taylor polynomials in $\epsilon$ around $\epsilon=0$, one solves the resulting equations for every order of the series separately. After neglecting the higher order terms in the viscosity ratios, one obtains evolution equations of $H_1$ and $H_2$. However, it is useful to also discuss the physical interpretation of the calculation, since the formal derivation leads to the same evolution equations as the following, physically more intuitive algorithm.

\begin{enumerate}[leftmargin=.5in]
\item[(i)\hspace{4pt}] For the first step one substitutes the liquid films with rigid walls. Their surfaces are not necessarily flat, but the gas velocity is assumed to be zero along the surfaces. Hence, during the first step one calculates the gas flow as if it were confined between two deformed walls. For this it is assumed that the volume flux of the gaseous medium is constant along the channel. To obtain the gas flow one first solves the equations assuming that $\Rey=0$, which gives the Stokes flow solution of the problem. Afterwards, the first-order perturbation of the inertial terms is calculated and is superposed to the previous solution. 
\item[(ii)\hspace{4pt}] After obtaining the flow field for the gaseous medium, one calculates the viscous stresses and the pressure gradient along the surfaces.
\item[(iii)\hspace{4pt}] The inertial terms in the liquid films are omitted, as they are $\textit{O}(1/\Mrat_1)^2$ and $\textit{O}(1/\Mrat_3)^2$. Hence, these layers are governed by the Stokes equation. Corresponding incompressible liquid films undergoing long-wavelength instabilities have been the subject of extensive research and are described by evolution equations for the film thickness (see \cite{oron_lubrication}). In this case they read

\begin{align}
\frac{\partial H_1}{\partial \tau}+\frac{\partial}{\partial X}\left[\frac{T{|}_{H_1}}{\Mrat_1}\frac{H_1^2}{2}-\frac{1}{\Mrat_1}\left(\frac{\partial P_2}{\partial X}{\Bigg |}_{H_1}-\frac{\epsilon}{Ca}\frac{\partial^3 H_1}{\partial X^3}\right)\frac{H_1^3}{3}\right]&=0,\label{eq:ev_eq_phys1}\\
\frac{\partial H_2}{\partial \tau}+\frac{\partial}{\partial X}\left[-\frac{T{|}_{1-H_2}}{\Mrat_3}\frac{H_2^2}{2}-\frac{1}{\Mrat_3}\left(\frac{\partial P_2}{\partial X}{\Bigg |}_{1-H_2}-\frac{\epsilon}{\Ca}\frac{\partial^3 H_2}{\partial X^3}\right)\frac{H_2^3}{3}\right]&=0.\label{eq:ev_eq_phys2}
\end{align}

The viscous stresses imposed by the gas flow at the two surfaces ($Y=H_1$ and $Y=1-H_2$) are expressed by $T=\partial V_X/\partial Y$.
\item[(iv)\hspace{4pt}] As the evolution equations are only dependent on the viscous stresses and the pressure gradients along the surface, the substitution of the results of step \text{(ii)} will give a self-contained differential equation system for $H_1$ and $H_2$, which is solved for one time increment.
\item[(v)\hspace{4pt}] Subsequently, one returns to the beginning of the algorithm, and the updated deformation of the liquid films will be used to calculate the new gas flow field according to (i).
\end{enumerate}

A detailed analysis of the limits of the assumptions introduced in this section is given in section \ref{sec:validification}.

After applying all assumptions and performing steps (i) and (ii), one finds that the pressure gradients and the viscous stresses at the surfaces are given by
\begin{align}
\frac{\partial P_2}{\partial X}{\Bigg |}_{H_1}=\frac{\partial P_2}{\partial X}{\Bigg |}_{1-H_2}&=-\frac{6 \left[70+9 \epsilon\Rey \left(\partial H_1/\partial X+\partial H_2/\partial X\right)\right]}{35 \left(1-H_1-H_2\right){}^3},\label{eq:press_approx}\\
T{|}_{H_1}=-T{|}_{1-H_2}&=\frac{6 \left[35+\epsilon\Rey \left(\partial H_1/\partial X+\partial H_2/\partial X\right)\right]}{35 \left(1-H_1-H_2\right){}^2}.\label{eq:stress_approx}
\end{align}

\begin{figure}
\centerline{\includegraphics[width=1\linewidth]{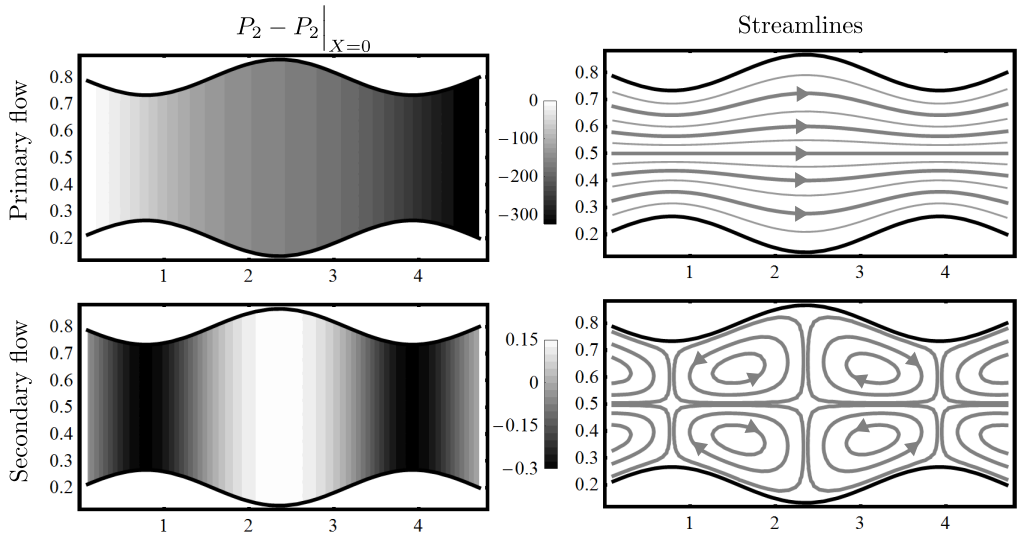}}\caption{Pressure drop and streamlines of an exemplary gas flow in a channel with symmetrically deformed walls. The Stokes flow (primary flow) is depicted in the first row, while the first-order correction arising from the non-zero inertial terms is shown in the second one. The full flow field corresponds to the superposition of the two results. ($\Rey=2$, $\epsilon=0.1$)}\label{fig:flow_example}
\end{figure}

An example of the flow field for $\Rey=2$ and $\epsilon=0.1$ in a symmetrically deformed channel is shown in figure \ref{fig:flow_example}. In the first row one can see the pressure drop and the streamlines of the Stokes flow while the inertial perturbation is depicted in the second row. The full flow field can be obtained by superposing the two results. In a later section and based on this illustration, the significance of the normal stresses to induce the instability will be compared to that of the shear stresses. Furthermore, one may ask how the film deformation affects the pressure gradient to maintain a constant volumetric flow rate. Based on equation \eqref{eq:press_approx}, one can calculate the pressure drop along the channel of length $L$. Assuming that the deformations at its ends are negligibly small yields
\vspace{-4pt}
\begin{equation}
P(L)-P(0)=-\int\limits_0^{L}{\frac{12}{(1-H_1-H_2)^3}dX}.
\end{equation}
\vspace{-8pt}

\noindent This implies that the terms scaling with the Reynolds number do not contribute to the pressure drop along the channel. Furthermore, the effect of the gas dragging the liquid films is not considered in the expression. The corresponding term contributes to the evolution equations only at higher orders of the viscosity ratios, which is beyond the scope of the current analysis. Assuming the system to be mirror-symmetrical and introducing the notation $H_1=H_2=H_0+\Delta H$, with $H_0$ being the initial uniform film thickness and the local deviation from that state is given by $\Delta H$, the integral takes the form 
\vspace{-6pt}
\begin{equation}
P(L)-P(0)=-\int\limits_0^{L}{\frac{12}{(1-2H_0)^3}+\frac{72 \Delta H}{(1-2H_0)^4}+\frac{288 \Delta H^2}{(1-2H_0)^5}+\frac{960 \Delta H^3}{(1-2H_0)^6}+...dX}.
\end{equation}

\noindent The linear term in $\Delta H$ drops out, since the total volume of the liquid films does not change during their evolution. Thus, if $|\Delta H/(1-2H_0)|\ll1$, the term scaling with $\Delta H^2$ defines the dominant effect of the deformations of the liquid surfaces. This term implies the pressure drop to be $1+O(\Delta H)^2\times24/(1-2H_0)^2$ times larger due to the instability of the interfaces.

Substituting the results for the pressure gradient and the viscous stresses leads to the following evolution equations:
\begin{align}
&\begin{aligned}
\frac{\partial H_1}{\partial \tau}=&-\frac{1}{\Mrat_1}\frac{\partial}{\partial X}{\Bigg\{}\frac{H_1^2(3+H_1-3H_2)}{(1-H_1-H_2)^3}+\epsilon \frac{H_1^3}{3\Ca_1}\frac{\partial^3 H_1}{\partial X^3}\\
+&\epsilon \Rey\frac{3H_1^2(1+5H_1-H_2)}{35(1-H_1-H_2)^3}\left(\frac{\partial H_1}{\partial X}+\frac{\partial H_2}{\partial X}\right){\Bigg\}},
\end{aligned}\label{eq:ev_eg_1}\\
&\begin{aligned}
\frac{\partial H_2}{\partial \tau}=&-\frac{1}{\Mrat_3}\frac{\partial}{\partial X}{\Bigg\{}\frac{H_2^2(3+H_2-3H_1)}{(1-H_1-H_2)^3}+\epsilon \frac{H_2^3}{3\Ca_3}\frac{\partial^3 H_2}{\partial X^3}\\
+&\epsilon \Rey\frac{3H_2^2(1+5H_2-H_1)}{35(1-H_1-H_2)^3}\left(\frac{\partial H_1}{\partial X}+\frac{\partial H_2}{\partial X}\right){\Bigg\}}.
\end{aligned}\label{eq:ev_eg_2}
\end{align}

A thorough analysis of the evolution equations is difficult, as the number of independent parameters in the equations is large. Therefore, all following sections of the paper will focus only on systems with liquid films of identical material properties. Formally, this implies $\Mrat_1=\Mrat_3\equiv \Mrat$ and $\Ca_1=\Ca_3\equiv \Ca$. At this point it is also useful to rescale the temporal and spatial coordinates as $\overline{\tau}=\tau\epsilon\sqrt[3]{\Ca}/\Mrat$ and $\overline{X}=\sqrt[3]{\Ca}X$. To avoid complicated notations, the overbars will be omitted in the following equations. The rescaled equations are given by

\begin{align}
&\begin{aligned}
\frac{\partial H_1}{\partial \tau}=&-\frac{\partial}{\partial X}{\Bigg\{}\frac{H_1^2(3+H_1-3H_2)}{\epsilon(1-H_1-H_2)^3}+\frac{H_1^3}{3}\frac{\partial^3 H_1}{\partial X^3}\\
+& \Rey\sqrt[3]{\Ca}\frac{3H_1^2(1+5H_1-H_2)}{35(1-H_1-H_2)^3}\left(\frac{\partial H_1}{\partial X}+\frac{\partial H_2}{\partial X}\right){\Bigg\}},
\end{aligned}\label{eq:ev_eg_sym_1}\\
&\begin{aligned}
\frac{\partial H_2}{\partial \tau}=&-\frac{\partial}{\partial X}{\Bigg\{}\frac{H_2^2(3+H_2-3H_1)}{\epsilon(1-H_1-H_2)^3}+\frac{H_2^3}{3}\frac{\partial^3 H_2}{\partial X^3}\\
+&\Rey\sqrt[3]{\Ca}\frac{3H_2^2(1+5H_2-H_1)}{35(1-H_1-H_2)^3}\left(\frac{\partial H_1}{\partial X}+\frac{\partial H_2}{\partial X}\right){\Bigg\}}.
\end{aligned}\label{eq:ev_eg_sym_2}
\end{align}

\section{Symmetrical systems}\label{sec:sym_sys}

This section discusses systems where the undisturbed liquid film thicknesses are equal, i.e. the undisturbed system is completely symmetrical with respect to the channel centre plane. The question arises whether or not the mirror symmetry breaks down during the evolution of the system. As a similar problem, \cite{hewitt_annular_flow} commented on the experimentally observable breakdown of axial symmetry in core-annular flows, supporting the significance of this question for planar flows.

Introducing the initial film thickness $H_0$, we denote the film deformations as $\Delta H_1=H_1-H_0$ and $\Delta H_2=H_2-H_0$. In the case of small deformations $\Delta H_1\equiv\epsilon \eta_1$, $\Delta H_2\equiv\epsilon \eta_2$, where $\textit{O}(\eta_1)=\textit{O}(\eta_2)=\textit{O}(1)$. Then one finds that - neglecting second or higher orders of $\epsilon$ - equations \eqref{eq:ev_eg_sym_1} and \eqref{eq:ev_eg_sym_2} transform to
\begin{align}
&\begin{aligned}
\frac{\partial \eta_1}{\partial \tau}=&-\frac{\partial}{\partial X}{\Bigg\{}\frac{6H_0\left[(1-H_0)\eta_1+H_0\eta_2\right]}{\epsilon(1-2H_0)^4}\\
+&\frac{3\left[(1-H_0)\eta_1+H_0\eta_2\right]}{(1-2H_0)^5}\left[(1+H_0(3-2H_0))\eta_1+H_0(3+2H_0)\eta_2\right]+\frac{H_0^3}{3}\frac{\partial^3 \eta_1}{\partial X^3}\\
+& \Rey\sqrt[3]{\Ca}\frac{3H_0^2(1+4H_0)}{35(1-2H_0)^3}\left(\frac{\partial \eta_1}{\partial X}+\frac{\partial \eta_2}{\partial X}\right){\Bigg\}},
\end{aligned}\label{eq_weak_nonlin1}\\
&\begin{aligned}
\frac{\partial \eta_2}{\partial \tau}=&-\frac{\partial}{\partial X}{\Bigg\{}\frac{6H_0\left[(1-H_0)\eta_2+H_0\eta_1\right]}{\epsilon(1-2H_0)^4}\\
+&\frac{3\left[(1-H_0)\eta_2+H_0\eta_1\right]}{(1-2H_0)^5}\left[(1+H_0(3-2H_0))\eta_2+H_0(3+2H_0)\eta_1\right]+\frac{H_0^3}{3}\frac{\partial^3 \eta_2}{\partial X^3}\\
+& \Rey\sqrt[3]{\Ca}\frac{3H_0^2(1+4H_0)}{35(1-2H_0)^3}\left(\frac{\partial \eta_1}{\partial X}+\frac{\partial \eta_2}{\partial X}\right){\Bigg\}}.
\end{aligned}\label{eq_weak_nonlin2}
\end{align}
The first and the second term in the curled brackets represent the stresses acting from the primary flow of the gas on the film. Hence, the governing equations remain non-linear in $\eta_1$ and $\eta_2$ even in this case. As it was argued by \cite{papaef_2013}, such a non-linearity can not be scaled out by a Galilei transformation. This restricts the validity of the linear stability analysis of the equations to a very narrow region around the equilibrium state.

\subsection{Linear stability analysis}\label{sec:lin_stab_anal}

Although equations \eqref{eq_weak_nonlin1} and \eqref{eq_weak_nonlin2} imply that the linearized equations only give a reliable approximation for the evolution of the interfaces for extremely small deformations, their analysis is helpful for the subsequent analysis of the weakly non-linear equations. Furthermore, the linear analysis also gives approximate values for the scaling parameters of the system.

Assuming that, instead of $\textit{O}(\eta_1)=\textit{O}(\eta_2)=\textit{O}(1)$ proposed for equations \eqref{eq_weak_nonlin1} and \eqref{eq_weak_nonlin2}, $\eta_1,\eta_2\ll\epsilon$, one can neglect the second-order terms in the deformations from equations \eqref{eq_weak_nonlin1} and \eqref{eq_weak_nonlin2}. The resulting system of linear partial differential equations can be analysed utilizing Fourier transformation. One finds the solution of the equations to be given by
\vspace{-4pt}
\begin{equation}\label{eq:lin_sol}
\begin{aligned}
&\begin{pmatrix}\eta_1\\ \eta_2\end{pmatrix}=\int\limits_{-\infty}^{\infty}{\left(A_+\bm{v_+}e^{\sigma_+\tau}+A_-\bm{v_-}e^{\sigma_-\tau}\right)e^{\iu kX}dk},\text{ where}\\
&\begin{aligned}
&\bm{v_+}=\begin{pmatrix}1\\1\end{pmatrix},&{\sigma_+=-\iu k \frac{6H_0}{\epsilon(1-2H_0)^4}-k^4\frac{H_0^3}{3}+k^2\Rey\sqrt[3]{\Ca}\frac{6}{35}\frac{H_0^2(1+4H_0)}{(1-2H_0)^3}},\\
&\bm{v_-}=\begin{pmatrix}1\\-1\end{pmatrix},&{\sigma_-=-\iu k \frac{6H_0}{\epsilon(1-2H_0)^3}-k^4\frac{H_0^3}{3}},
\end{aligned}
\end{aligned}
\end{equation}

\noindent with $k$ being the non-dimensional wave number and $A_{\pm}(k)$ being the Fourier amplitudes. From this it is apparent that the deformations of the interfaces are the superposition of a symmetrical instability, $\bm{v_+}$, and a linearly independent antisymmetrical one, $\bm{v_-}$. As $\Imag(\sigma_\pm)\propto k$, the linear interfacial waves do not disperse. Furthermore, $\Real(\sigma_-)<0$ for all wavenumbers, while $\Real(\sigma_+)>0$ for sufficiently small wavenumbers. Consequently, the interfaces are always linearly unstable for the symmetrical deformations, while the antisymmetrical ones are linearly stable.

On the one hand, the characteristic (i.e. dominant) wavenumber of the deformations can be assumed to correspond to the wavenumber maximizing $\sigma_+$ (i.e. the fastest growing wavenumber), which is given by
\begin{equation}\label{eq:nondim_wavenum}
k_{\text{char}}=\left[{\frac{9}{35}\Rey\sqrt[3]{\Ca}\frac{1+4H_0}{H_0(1-2H_0)^3}}\right]^{1/2}.
\end{equation}

\noindent Hence, owing to the negligible effect of gravity, the films are always inherently unstable to disturbances of sufficiently small  wavenumbers as long as $\Rey>0$. 

On the other hand, according to section \ref{sec:gov_eq}, the non-dimensional $X$ coordinate is given by $X=x_1\sqrt[3]{\Ca}/\lambda_{\text{char}}$, where $\lambda_{\text{char}}$ is the characteristic dimensional wavelength of the deformations. Thus, consistent scaling requires that $k_{\text{char}}=2\pi/\sqrt[3]{Ca}$. The equivalence of this wavenumber to the one from equation \eqref{eq:nondim_wavenum} provides an expression for $\Ca$ given by
\vspace{-4pt}
\begin{equation}
\Ca=\frac{140\pi^2H_0(1-2H_0)^3}{9\Rey(1+4H_0)}.\label{eq:ca_cons}
\end{equation}
With the definition of the Capillary number one may also compute $\epsilon$ to be given by
\begin{equation}\label{eq:eps_form}
\epsilon=\sqrt{\frac{9\mu_2Q\Rey(1+4H_0)}{140\pi^2d\gamma H_0(1-2H_0)^3}}=\sqrt{\frac{9\rho_2Q^2(1+4H_0)}{140\pi^2d\gamma H_0(1-2H_0)^3}}.
\end{equation}

\noindent The neutrally stable wavenumber is obtained from the condition $\text{Re}(\sigma_+)=0$. According to equation \eqref{eq:lin_sol}, this condition is satisfied at $k_{\text{max}}=\sqrt{2}k_{\text{char}}$.

\subsubsection{Physical interpretation of the results}

Firstly, one may utilize the results of the linear stability analysis to judge the relative influence of the pressure gradient and the viscous stresses, respectively, on the instability. For this purpose, one assumes that both effects may be considered separately from one another. Without the presence of viscous stresses along the interface, the system would still be unstable, but with the characteristic wavenumber $k_{\text{char}}^\text{P}$. This instability emerges due to a mechanism analogous to the Venturi effect, i.e. it is caused by the fluctuation in the dynamic pressure of the gas flow triggered by the widening or narrowing of the gas layer with changing liquid-layer thicknesses. This is depicted in the bottom-left graph of figure \ref{fig:flow_example}. Similarly, if the pressure gradient is omitted from the evolution equations \eqref{eq:ev_eq_phys1} and \eqref{eq:ev_eq_phys2}, one obtains $k_{\text{char}}^\text{T}$ for the characteristic wavenumber. The viscous instability is caused by a "sweeping" mechanism driven by the counter-rotating vortices of the secondary air flow before and after an elevation of the liquid-gas interface, as it is shown in the bottom-right graph of figure \ref{fig:flow_example}. A straightforward calculation leads to
\vspace{-2pt}
\begin{align}
k_{\text{char}}^\text{P}&=\left[\frac{54}{35}\Rey\sqrt[3]{\Ca}\frac{1}{(1-2H_0)^3}\right]^{1/2},\\
k_{\text{char}}^\text{T}&=\left[\frac{9}{35}\Rey\sqrt[3]{\Ca}\frac{1}{(1-2H_0)^2H_0}\right]^{1/2}.
\end{align}

\noindent The obtained wavenumbers can be related to the original one with the expression $k_{\text{char}}^2=(k_{\text{char}}^\text{P})^2+(k_{\text{char}}^\text{T})^2$. Moreover, as long as $H_0<0.125$ ($H_0>0.125$), $k_{\text{char}}^\text{T}>k_{\text{char}}^\text{P}$ ($k_{\text{char}}^\text{T}<k_{\text{char}}^\text{P}$). This suggests that at small film thicknesses the interfaces are mainly deformed by viscous stresses, while for larger film thicknesses the instabilities are driven by pressure gradients. 

Secondly, one may be interested in the effect the coupling between the two interfaces has on the emergence of the patterns. That is, similarly as treated by \cite{coupled_sos}, one can analyse how the properties of the instabilities of the two coupled interfaces differ from those of the instability that would appear if the deformation of one layer would not affect the other interface. The mathematical implementation of the latter scenario corresponds to setting $\eta_2=0$ in equation \eqref{eq_weak_nonlin1} and $\eta_1=0$ in equation \eqref{eq_weak_nonlin2}. Performing a linear stability analysis indicates that the characteristic wavenumber of such a configuration is given by
\vspace{-4pt}
\begin{equation}\label{eq:uncoupled_wavenum}
k_{\text{char}}'=\left[{\frac{9}{70}\Rey\sqrt[3]{\Ca}\frac{1+4H_0}{H_0(1-2H_0)^3}}\right]^{1/2}.
\end{equation}

\noindent Hence, the characteristic wavenumber is $\sqrt{2}$ times smaller than for the coupled system. Furthermore, its corresponding growthrate is one fourth of the growthrate of the coupled system at $k=k_{\text{char}}$. Hence, the coupling between the interfaces further destabilizes the system. 

To put these expressions for the wavenumbers into a bigger picture, the (dimensional) characteristic wavenumber of the classical inviscid (and uncoupled) Kelvin-Helmholtz instability of a deep, quiescent liquid layer subjected to a cross flow of a gas is given by \citep{chandrasekhar_hydrodynamic}
\begin{equation}
k_{\text{char,dim}}^{\text{KH}}=\frac{2\rho_1\rho_2u^2}{3(\rho_1+\rho_2)\gamma}=\frac{2\rho_1\rho_2\{Q/[d(1-2H_0)]\}^2}{3(\rho_1+\rho_2)\gamma},
\end{equation}

\noindent where gravity was neglected for consistency, while $\rho_1$ and $\rho_2$ denote the fluid densities. The mean gas velocity was approximated by its volumetric flux according to $u=Q/[d(1-2H_0)]$. After non-dimensionalisation one obtains
\vspace{-2pt}
\begin{equation}
k_{\text{char}}^{\text{KH}}=\frac{\rho_1\sqrt{\rho_2}Q}{(\rho_1+\rho_2)\sqrt[3]{Ca}}\sqrt{\frac{560\pi^2 H_0}{81\gamma d(1+4H_0)(1-2H_0)}}.
\end{equation}

\noindent For instance, for an air flow between silicone oil films, with a channel width of $d=200\text{\,}\mu\text{m}$, film thickness $H_0=0.1$ and mean air velocity of $1\text{\,m\,s}^{\text{-}1}$, the ratio of the characteristic wavenumbers is $k_{\text{char}}^{\text{KH}}/k_{\text{char}}'\approx0.025$. Thus, the classical Kelvin-Helmholtz instability induces deformations at the interface of a deep liquid layer, which exhibit considerably longer wavelengths than the ones obtained in this paper for very thin films. The material properties used for the approximation were taken from the first column of table \ref{tab:mat_prop}.

\subsubsection{Validity range of the assumptions}\label{sec:validification}

Regarding the requirement that $\epsilon^2\ll1$, the permittable volumetric fluxes as a function of $H_0$ and the material properties are readily given by equation \eqref{eq:eps_form}.

With respect to $\epsilon \Rey$, the corresponding first-order perturbation approach is readily justified for small $\epsilon \Rey$. However, one expects the calculation to be sufficiently accurate as long as the convection-related corrections to the pressure gradient and the shear stresses in the gas layer, expressed by the terms in \eqref{eq:press_approx} and \eqref{eq:stress_approx} multiplied by $\epsilon Re$, remain small. The results of the linear stability analysis can be used to show that equations \eqref{eq:ev_eg_1} and \eqref{eq:ev_eg_2} remain valid also at larger values of $\epsilon \Rey$.  To show this, one assumes $\text{O}(\partial H_1/\partial X)\approx\text{O}(\partial H_2/\partial X)\approx\Delta Hk_{\text{char}}$. According to equations \eqref{eq:press_approx} and \eqref{eq:stress_approx}, if these corrections are small, then the perturbation approach for the inertial flow is justified even for larger values of $\epsilon Re$. As the relative effect of the inertial flow is approximately $4.5$ times larger for the pressure gradient than for the viscous stresses, it is sufficient to consider only the former term. For an antisymmetrical deformation, the correction vanishes completely, while for a symmetrical deformation, the contribution of inertia to the pressure gradient is given by
\vspace{-4pt}
\begin{equation}\label{eq:eff_inflow}
\Delta P'\approx\frac{9}{35}\epsilon \Rey k_{\text{char}}\Delta H.
\end{equation}

\noindent Thus, as long as $(\Delta P')^2\ll1$ the perturbation approach is valid also at larger values of $\epsilon Re$. To estimate the value of $\Delta H$ one may assume that it will be of the same order of magnitude as $H_0$. Hence, by substituting $\Delta H$ with $H_0$, one may estimate the value of $\Delta P'$ directly, utilizing only the properties of the initial setup of the system.

With respect to the first-order expansion in $1/\Mrat$, estimating the error introduced by neglecting terms of higher order than $1/M$ from the evolution equations requires a more careful approach, as $\Mrat$ is absent from the expressions \eqref{eq:ev_eg_sym_1} and \eqref{eq:ev_eg_sym_2}. Although \cite{aturkar_lin_multi} and \cite{renardy_triple_layer} discussed the linear behaviour of multi-layer systems of arbitrary viscosities, they did not limit their description to interfaces undergoing long-wavelength instabilities. Therefore, the equations arising from their calculations are more complicated and are only accessible by numerical simulations. Thus, in order to perform a detailed analytical description for arbitrary viscosities but within the framework of the long-wavelength approximation, it is necessary to re-derive the linearized equations, starting from \eqref{eq:nondim_bulk_ns} - \eqref{eq:cont_int}, as will be described in the following.

\begin{figure}
\centerline{\includegraphics[width=1\linewidth]{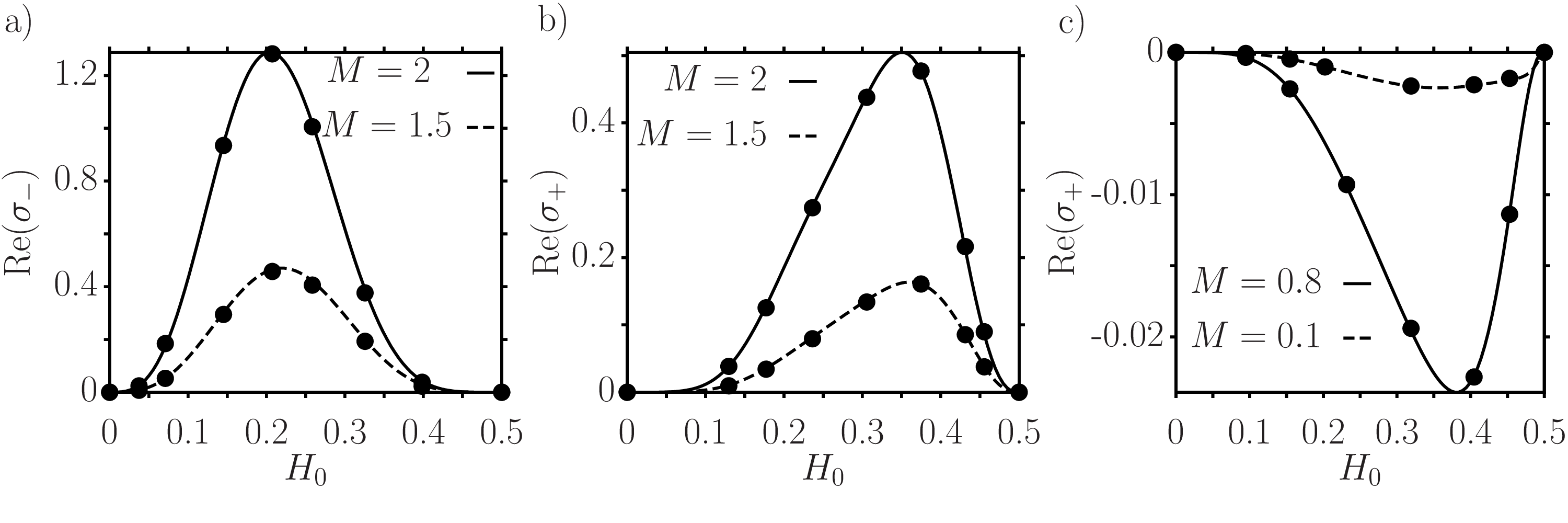}}\caption{Comparison of the growthrates of the symmetrical $\sigma_+$ and the antisymmetrical $\sigma_-$ mode, according to expressions \eqref{eq:general_visc1} and \eqref{eq:general_visc2}, with the numerical results of \cite{renardy_triple_layer}. The dots represent the numerical results, which are taken from figures 2, 4 and 3, respectively, of the referred paper.  The lines correspond to the analytical formulas. ($\mu_1=\mu_3=1.56605\times10^{\text{-}6}\text{\,Pa\,s}$, $d=200\text{\,}\mu \text{m}$, $\rho_1=\rho_2=\rho_3=1\text{\,kg\,m}^{\text{-}3}$, $\partial p/\partial x=19.62\text{\,\,Pa\,m}^{\text{-}1}$, $\gamma\rightarrow 0$ (see appendix \ref{app_growth} for explanation), $k=0.01/H_0$)}\label{fig:renardy}
\end{figure}

Within the long-wavelength approximation for systems featuring arbitrary viscosity ratios, a feasible solution approach is to simultaneously linearize the full Navier-Stokes equations \eqref{eq:nondim_bulk_ns}-\eqref{eq:cont_int} in terms of both $\epsilon Re$ and the interfacial deformations, $\eta_1$ and $\eta_2$. For asymmetrical systems the calculations are overly complicated, but for symmetrical systems the equations are solvable. Nonetheless, the formulas emerging from this direct approach would be still too extensive to handle. However, the mirror symmetry of the system suggests that the solution of the linearized problem should be separable into a symmetrical and an independent antisymmetrical instability mode, even if the intermediate layer has an arbitrary viscosity. Formally, the linear solution takes the same form as equation \eqref{eq:lin_sol}, where the eigenvectors remain unchanged and the difference only arises in the expression for the eigenvalues. Given the linear independency of the eigenvectors one may calculate the growthrates for the symmetrical and the antisymmetrical modes separately. To this end, one first assumes that the small interfacial deformation of the interfaces is symmetrical, which corresponds to eigenvector $\bm{v_+}$. This leads to a single evolution equation for the deformations $\eta_1=\eta_2$. Examining its Fourier transform will give the value of $\text{Re}(\sigma_+)$. Similarly, one obtains $\text{Re}(\sigma_-)$ by solving the linearized problem while assuming an antisymmetrical deformation of the interfaces. The calculations are straightforward, but cumbersome and they were performed with the \textit{Mathematica}$\textsuperscript{\textregistered}$ software. A comparison of the formulas obtained with this method and the numerical results of \cite{renardy_triple_layer} is presented in figure \ref{fig:renardy}. The graphs show a good agreement between the two approaches. The complete expressions for the growthrates at arbitrary viscosity ratios and a more detailed discussion of the governing parameters of figure \ref{fig:renardy} are given in appendix \ref{app_growth}. 

With these results, it is useful to calculate the relative error between the exact (linear) growthrates for arbitrary viscosities and the ones obtained for the simplified problem, where the middle layer has a much smaller viscosity than the liquid films. Denoting the exact growthrates by $\text{Re}(\sigma_{+}^{e})$ and $\text{Re}(\sigma_{-}^{e})$, the relative error of the simplification is defined by 
\vspace{-4pt}
\begin{equation}\label{eq:err_visc}
\Delta\sigma_\pm^2=\frac{\int\limits_0^{k_\text{max}}{\left[\text{Re}(\sigma_{\pm}^{ e})-\text{Re}(\sigma_{\pm })\right]^2dk}}{\int\limits_0^{k_\text{max}}{\left[\text{Re}(\sigma_{\pm}^{ e})\right]^2dk}}.
\end{equation}

\begin{figure}
\centerline{\includegraphics[width=0.8\linewidth]{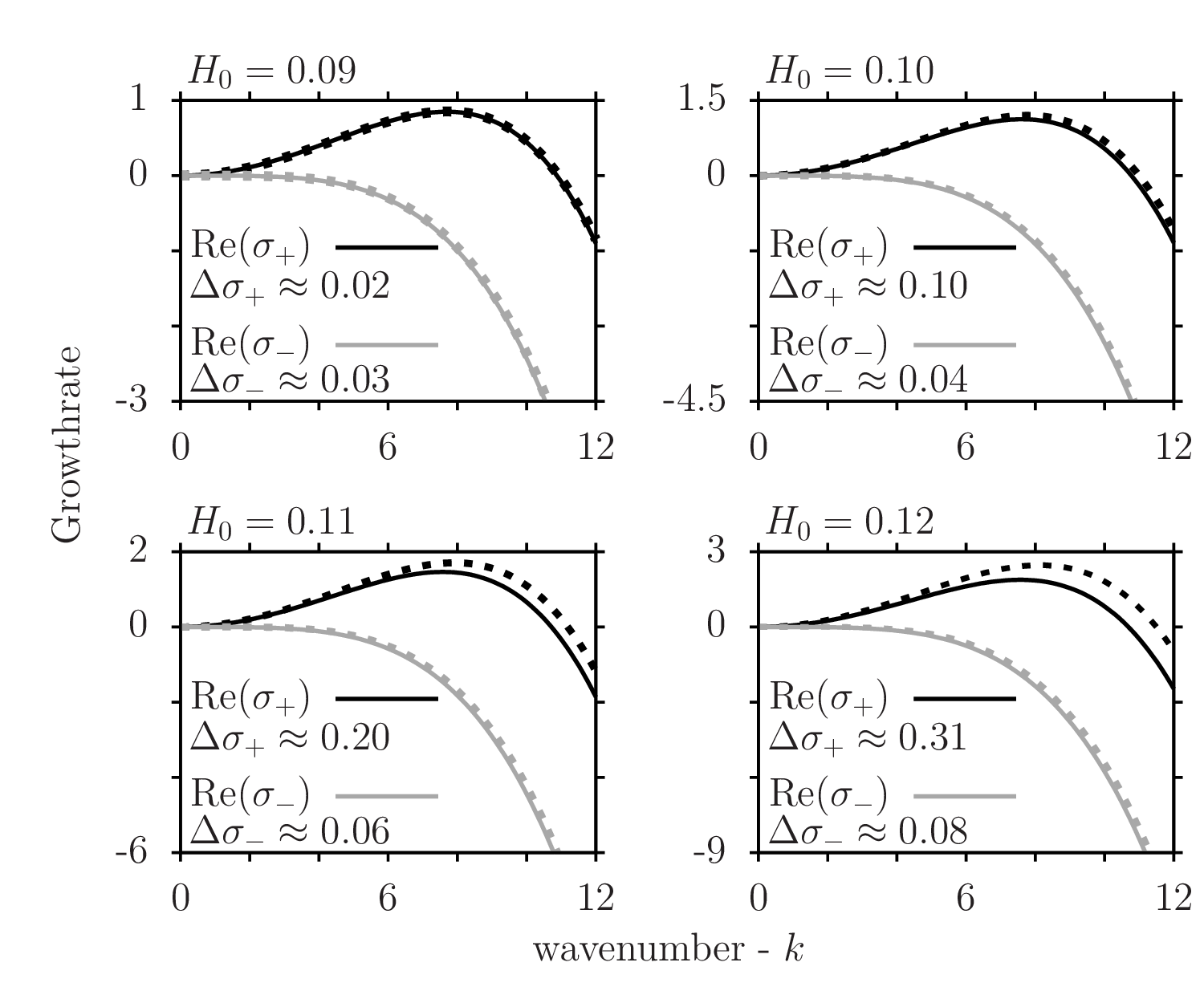}}
\caption{The real part of the eigenvalues, $\sigma_+$ and $\sigma_-$, for the linearized water-air system as a function of the wavenumber. The continuous lines indicate the results obtained under the assumption of semi-rigidness of the films ($M\gg1$), while the exact eigenvalues for the actual value of $\Mrat$ are plotted with dashed lines. The initial thickness $H_0 = h_0/d$ of the symmetrical liquid films is varied for each plot, as indicated ($d=200\text{\,}\mu \text{m}$, mean air velocity $1\text{\,m\,s}^{\text{-}1}$, $\Rey\approx 10$, $\Ca\approx 0.56$).}\label{fig:growthrate}
\end{figure}

\noindent The linearly neutrally stable wavenumber ($k_\text{max}=\sqrt{2}k_\text{char}$ with $k_\text{char}$ defined by \eqref{eq:nondim_wavenum}) is taken as the limit of the integrals, as one is mainly interested in the regime of unstable wavenumbers ($k<k_\text{max}$), for which $\text{Re}(\sigma_+)\geq0$. After substituting the expressions for the growthrates into \eqref{eq:err_visc}, one finds that $\Delta\sigma_\pm^2$ are independent of the Reynolds and the Capillary number. This is an important property, as it shows that even though the growthrates are functions of $\Rey\sqrt[3]{\Ca}$, $\Delta\sigma_\pm$ is only a function of the viscosity and density ratios and $H_0$. An example of the results for a water-air system is depicted in figure \ref{fig:growthrate}. Values for the material properties at $25 \text{\,}^\circ\text{C}$ and $1\text{\,bar}$ used to generate this figure are summarized in table \ref{tab:mat_prop}.

\begin{table}
  \centering
    \begin{tabular}{c l  l  l}
     \multicolumn{3}{c}{$T=25\text{\,}^\circ\text{C}$} \\[-5pt]
     \hline\\[-13pt]
  						& Parameter & Value\\[3pt]
	   Air   		&	Density 						& $1.161 {\text{ kg\,}}{\text{m}^{\text{-}3}}$\\
							& Viscosity 					& $18.54 \times 10^{-6} \text{ Pa\,s}$\\ 
			Water		&	Density 						& $997.05 {\text{ kg\,}}{\text{m}^{\text{-}3}}$\\
		 	 				& Surface tension 		& $71.99 \times10^{-3} {\text{ N}\,}{\text{m}^{\text{-}1}}$\\
			 				& Viscosity 					& $8.9 \times 10^{-4} \text{ Pa\,s}$\\
    \end{tabular}
  \begin{tabular}{c l  l  l}
  \multicolumn{3}{c}{$T=50\text{\,}^\circ\text{C}$}\\[-5pt]
  \hline\\[-13pt]
  			& Parameter & Value\\[3pt]
	   Air   		&	Density 						& $1.089 {\text{ kg}}\text{\,}{\text{m}}^{\text{-}3}$\\
							& Viscosity 					& $19.49 \times 10^{\text{-}6} \text{ Pa\,s}$\\ 
			Oil			&	Density 						& $940 {\text{ kg}}\text{\,}{\text{m}^{\text{-}3}}$\\
		 	 				& Surface tension 		& $19 \times10^{\text{-}3} {\text{ N}}\text{\,}{\text{m}^\text{-1}}$\\
			 				& Viscosity 					& $9.59 \times 10^{-3} \text{ Pa\,s}$\\
    \end{tabular}\caption{Material properties of air, water \citep{crc_handbook} and $10\text{\,}\text{cSt}$ silicone oil \citep{vanhook_lw_exp_theo} for different temperatures at $1\text{\,bar}$ pressure.}
    \label{tab:mat_prop}
    \end{table}

Figure \ref{fig:growthrate} confirms intuition in the sense that the assumption of semi-rigidness becomes less accurate with increasing the blockage of the total channel width by the liquid films. For the examples shown this is particularly noticeable for $H_0 \geq 0.11$. The approximations estimating the accuracy of the assumption of semi-rigidness are only valid for the equations linearized in $\eta_1$ and $\eta_2$. However, one does not expect the order of the relative error to change considerably for the non-linear equations, at least as long as the air gap between the liquid films does not become too small. This can be explained by the linearity of the viscous stress term in the Navier-Stokes equation.

\subsection{Weakly non-linear analysis}\label{sec:weak_nonlin}

As the eigenvectors of the linear equations $\bm{v_+}$ and $\bm{v_-}$ form a complete set for the function space of possible solutions, one can assume without loss of generality that the solutions of the weakly non-linear equations take the form
\vspace{-2pt}
\begin{equation}\label{eq:general_sol}
\begin{aligned}
&\begin{pmatrix}\eta_1\\ \eta_2\end{pmatrix}=\int\limits_{-\infty}^{\infty}{\left(A_+(k,\tau)\bm{v_+}e^{\sigma_+\tau}+A_-(k,\tau)\bm{v_-}e^{\sigma_-\tau}\right)e^{\iu kX}dk}.\\
\end{aligned}
\end{equation}

\noindent This expression is formally identical to the one presented in equation \eqref{eq:lin_sol}, except that here the Fourier coefficients are not only functions of the wavenumber $k$ but are also time-dependent. The growthrates $\sigma_\pm$ are still defined by the same expressions as in equation \eqref{eq:lin_sol}. After substituting this integral into equations \eqref{eq_weak_nonlin1} and \eqref{eq_weak_nonlin2} one obtains the relationships between the Fourier coefficients to be given by

\begin{equation}\label{eq:amp_ev}
\begin{aligned}
\frac{\partial A_{+}}{\partial \tau}e^{\sigma_+ \tau}+\iu k{\Big[}\frac{3(1+6H_0)}{(1-2H_0)^5}\left(A_+e^{\sigma_+\tau}\right)\underset{k}*\left(A_+e^{\sigma_+\tau}\right)&\\
+\frac{3(1+2H_0)}{(1-2H_0)^3}\left(A_-e^{\sigma_-\tau}\right)\underset{k}*\left(A_-e^{\sigma_-\tau}\right){\Big]}&=0,\\
\frac{\partial A_{-}}{\partial \tau}e^{\sigma_- \tau}+\iu k\frac{6(1+4H_0)}{(1-2H_0)^4}(A_+e^{\sigma_+\tau})\underset{k}*(A_-e^{\sigma_-\tau})&=0,\\
\text{where } f\underset{k}*g:=\int\limits_{-\infty}^\infty{f(k-k')g(k')dk'}&
\end{aligned}
\end{equation}

\noindent is the convolution of the functions with respect to the wavenumber. The derivation of these equations is discussed in detail in appendix \ref{ampl_eq}. For the calculations and the discussion of the actual film deformations it is useful to introduce the variables
\vspace{-2pt}
\begin{equation}
\zeta_+(X)=\int\limits_{-\infty}^{\infty}{A_+\expf[{\Real(\sigma_+)\tau+\iu k X}]dk} \quad \quad \quad \zeta_-(X)=\int\limits_{-\infty}^{\infty}{A_-\expf[\Real(\sigma_-)\tau+\iu k X]dk}.
\end{equation}
\vspace{-10pt}

\noindent Since in the definition of $\zeta_-$ and $\zeta_+$ the imaginary parts of the growthrates were omitted, one needs to consider them as a shift along the $X$ coordinate. Thus, the film deformations are given by
\begin{align}
\eta_1(X)&=\zeta_+(X-6H_0/\left[\epsilon(1-2H_0)^4\right]\tau)+\zeta_-(X-6H_0/\left[\epsilon(1-2H_0)^3\right]\tau),\\ \eta_2(X)&=\zeta_+(X-6H_0/\left[\epsilon(1-2H_0)^4\right]\tau)-\zeta_-(X-6H_0/\left[\epsilon(1-2H_0)^3\right]\tau),
\end{align}

 \noindent i.e. $\zeta_+$ and $\zeta_-$ define the magnitude of the symmetrical and the asymmetrical deformations at a point along the horizontal coordinate. An inverse Fourier transformation of equation \eqref{eq:amp_ev} results in the partial differential equations

\begin{align}
&\begin{aligned}
0=\frac{\partial \zeta_+}{\partial \tau}&+\frac{H_0^3}{3}\frac{\partial^4 \zeta_+}{\partial X^4}+ \Rey \sqrt[3]{\Ca}\frac{6}{35}\frac{H_0^2(1+4H_0)}{(1-2H_0)^3}\frac{\partial^2 \zeta_+}{\partial X^2}\\
&+\frac{3(1+6H_0)}{(1-2H_0)^5}\frac{\partial\zeta_+^2}{\partial X}+\frac{3(1+2H_0)}{(1-2H_0)^3}\frac{\partial\zeta_-^2}{\partial X}{\Bigg |}_{X+\frac{12H_0^2}{\epsilon(1-2H_0)^4}\tau},
\end{aligned}\label{eq:wn_zetap}\\
&\begin{aligned}
0=\frac{\partial \zeta_-}{\partial \tau}&+\frac{H_0^3}{3}\frac{\partial^4 \zeta_-}{\partial X^4}+ \frac{6(1+4H_0)}{(1-2H_0)^4}\frac{\partial\left[\zeta_-\left(\zeta_+{\Big |}_{X-\frac{12H_0^2}{\epsilon(1-2H_0)^4}\tau}\right)\right]}{\partial X}.
\end{aligned}\label{eq:wn_zetam}
\end{align}

\noindent The subscript $X\pm12H_0^2/[\epsilon(1-2H_0)^4]\tau$ represents the differences in the phase velocities of the symmetrical and the antisymmetrical mode. It indicates that one evaluates the corresponding function at a shifted value of $X$.

The equations show that while the (weakly non-linear) symmetrical deformation is affected by the inertial flow, the antisymmetrical mode is not directly influenced by it. Furthermore, both equations consist of local terms and a non-local coupling term between the instability modes. To describe this coupling, it is useful to transform the equations. Similarly to the approach used to analyse the Kuramoto-Sivashinsky equations \citep{chang_complex_wave}, one multiplies the equations \eqref{eq:wn_zetap} and \eqref{eq:wn_zetam} with $\zeta_+$ and $\zeta_-$, respectively, and integrates the resulting equations along $-\infty<X<\infty$. Assuming that for every $\tau$ the deformations and their derivatives vanish as $X\rightarrow \pm\infty$, (i.e. the interfaces are pinned at $X\rightarrow \pm\infty$ with a contact angle of $90^\circ$) one finds that
\vspace{-2pt}
\begin{align}
\begin{aligned}
\frac{\partial}{\partial \tau}\left<\zeta_+^2\right>-\Rey\sqrt[3]{\Ca} \frac{12}{35}\frac{H_{0}^2(1+4H_{0})}{(1-2H_{0})^3}\left<\left(\frac{\partial \zeta_+}{\partial X}\right)^2\right>+\frac{2H_0^3}{3}\left<\left(\frac{\partial^2 \zeta_+}{\partial X^2}\right)^2\right>&\\
-\frac{6(1+2H_0)}{(1-2H_0)^3}{\text{corr}}\left(\frac{\partial\zeta_+}{\partial X},\zeta_-^2\right){\Bigg |}_{\textstyle \frac{12H_0^2}{\epsilon(1-2H_0)^4}\tau}&=0,
\end{aligned}\label{eq:nonlin_av1}&\\
\begin{aligned}
\frac{\partial}{\partial \tau}\left<\zeta_-^2\right>+ \frac{2H_0^3}{3}\left<\left(\frac{\partial^2 \zeta_-}{\partial X^2}\right)^2\right>+\frac{6(1+4H_0)}{(1-2H_0)^4}{\text{corr}}\left(\frac{\partial\zeta_+}{\partial X},\zeta_-^2\right){\Bigg |}_{\textstyle \frac{12H_0^2}{\epsilon(1-2H_0)^4}\tau}=0,
\end{aligned}\label{eq:nonlin_av2}&\\
\begin{aligned}
\text{where } {\text{corr}}\left(\frac{\partial\zeta_+}{\partial X},\zeta_-^2\right){\Bigg |}_{\Delta X}=\int\limits_{-\infty}^{\infty}
{\frac{\partial\zeta_+}{\partial X}{\Bigg |}_{X}\zeta_-^2{\Bigg |}_{X+\Delta X}dX}.
\end{aligned}&\nonumber
\end{align} 

\noindent The equations remain the same if - instead of the condition of pinned interfaces - periodic boundary conditions are applied at the domain boundaries. The operator $\left<f(X)\right>$ represents the integration of $f(X)$ on $[-\infty,\infty]$. The quantities $\left<\zeta_+^2\right>$ and $\left<\zeta_-^2\right>$ characterize the magnitude of the symmetrical and the antisymmetrical instability mode, respectively, over the whole length of the interface. As $\left<\left({\partial \zeta_+}/{\partial X}\right)^2\right>$ is non-negative, the perturbation due to the inertial term always leads to an increase of the symmetrical deformations. Similarly, $\left<\left({\partial^2 \zeta_+}/{\partial X^2}\right)^2\right>$ is always non-negative as well, so that capillarity always decreases the deformation amplitudes. Therefore, the interfaces are not susceptible to the Majda-Pego instability. The absence of such instabilities in the system is discussed in more detail in appendix \ref{kin_inst}.

As summarized in appendix \ref{simres}, from numerical simulations it was found that the antisymmetrical mode is always dampened out. Intuitively, this can be understood as follows: equations \eqref{eq:nonlin_av1} and \eqref{eq:nonlin_av2} suggest that the antisymmetrical instability mode can only be destabilized if the correlation function between the antisymmetrical and the symmetrical modes is negative. In this case equation \eqref{eq:nonlin_av1} indicates that $\left<\zeta_+^2\right>$ grows slower than for configurations for which the correlation is positive. In turn, for a positive correlation function the symmetrical deformation further dampens the antisymmetrical instability. Thus, one expects the symmetrical instability mode to have a generally stabilizing effect on the antisymmetrical one. Moreover as $\tau\rightarrow\infty$, the spatial distance over which the correlation is evaluated also tends to infinity. Although dissipative systems are known to exhibit long-range coherent behaviour \citep{nicolis_self_org}, the correlation function still needs to vanish in physically realistic setups as the shift between $\zeta_+$ and $\zeta_-$ tends to infinity. Therefore, after sufficiently long time, the coupling between the instability modes will have a negligible effect on the film evolution. Both arguments, supported by numerical tests, suggest that the antisymmetrical instability mode decays during the long-time evolution of the interfaces, which can be formally expressed by $\left<\zeta_-^2\right>\rightarrow0$, thus $\zeta_-\rightarrow 0$ at every point. Hence, as long as the symmetrical mode is unstable, one may assume that as time proceeds $\zeta_-\ll\zeta_+$. From this and equation \eqref{eq:wn_zetap} one can deduce that the physically relevant symmetrical instability mode will evolve according to the Kuramoto-Sivashinsky equation given by
\vspace{-2pt}
\begin{equation}\label{eq:zeta_ks}
0=\frac{\partial \zeta_+}{\partial \tau}+\frac{H_0^3}{3}\frac{\partial^4 \zeta_+}{\partial X^4}+ \Rey \sqrt[3]{\Ca}\frac{6}{35}\frac{H_0^2(1+4H_0)}{(1-2H_0)^3}\frac{\partial^2 \zeta_+}{\partial X^2}+\frac{3(1+6H_0)}{(1-2H_0)^5}\frac{\partial\zeta_+^2}{\partial X}.
\end{equation}

\noindent Non-linear partial differential equations of this type have been first derived by \cite{kuramoto} for reaction-diffusion systems and by \cite{sivashinsky} for liquid-film instabilities. They have been the subject of extensive analytical (\cite{chang_complex_wave}, \cite{kudryashov_exact_sol}) and numerical studies (\cite{cvitanovic_phase_space}, \cite{kevrekedis_comp}). Therefore, the further analysis of these equations will not be discussed here in detail, the reader is referred to the existing literature on the topic. 

After calculating $\zeta_+$ with \eqref{eq:zeta_ks} (under the assumption that the coupling to the antisymmetrical mode is negligibly small), one can solve equation \eqref{eq:wn_zetam} independently to obtain $\zeta_-$. However, it is noted that even when $\zeta_-$ is not negligibly small compared to $\zeta_+$, equations \eqref{eq:wn_zetap} and \eqref{eq:wn_zetam} are still reducible to a single evolution equation. The derivation of this quite complicated equation is discussed in appendix \ref{reduction}. 

According to \cite{hyman_kuramoto_limit}, equations of the form
\vspace{-2pt}
\begin{equation}\label{eq:hyman}
\frac{\partial u}{\partial t}+\nu\frac{\partial^4 u}{\partial x^4}+\frac{\partial^2 u}{\partial x^2}+\frac{1}{2}\frac{\partial u^2}{\partial x}=0
\end{equation}

\noindent should have a bounded solution in the long-time limit, where an upper bound scales like $\nu^{-1/4}$. After rescaling of equation \eqref{eq:zeta_ks} to take the same form as \eqref{eq:hyman}, this condition transforms to
\vspace{-2pt}
\begin{equation}\label{eq:kura_lim}
\lim\limits_{\tau\rightarrow\infty}\text{sup}(\zeta_+)\le \text{const}{\Bigg [}\frac{(1+6H_0)^2}{H_0^3(1-2H_0)(1+4H_0)^3CaRe^3}{\Bigg ]}^{-1/4}.
\end{equation}

\noindent Therefore, the amplitudes of the interfacial deformation can not increase indefinitely but should be stabilized by the non-linear terms in the evolution equation. This property is consistent with the physical intuition. However, due to the unknown constant on the right-hand side of the expression, giving an exact upper limit for the deformation amplitudes is not possible.

As mentioned above, in order to examine the long-term damping of the antisymmetrical instability mode, a number of simulations of the full evolution equations \eqref{eq:ev_eg_sym_1} and \eqref{eq:ev_eg_sym_2} were conducted. The finite-element method with quadratic (Lagrangian) shape functions was applied for solving the equations. The calculations were performed with \cite{comsol} using the Matlab Livelink environment. The non-dimensional parameters were varied according to $H_0\in[0.01,0.3]$, $\epsilon\in[0.01,0.5]$, $\Rey\in[1,50]$, $\Ca\in[0.125,5]$. The simulation results are summarized in appendix \ref{simres}. They all indicate that the antisymmetrical mode is quickly damped out. Thus, for symmetrical systems, the films fully synchronize in phase and amplitude.

\begin{figure}
\centerline{\includegraphics[width=\linewidth]{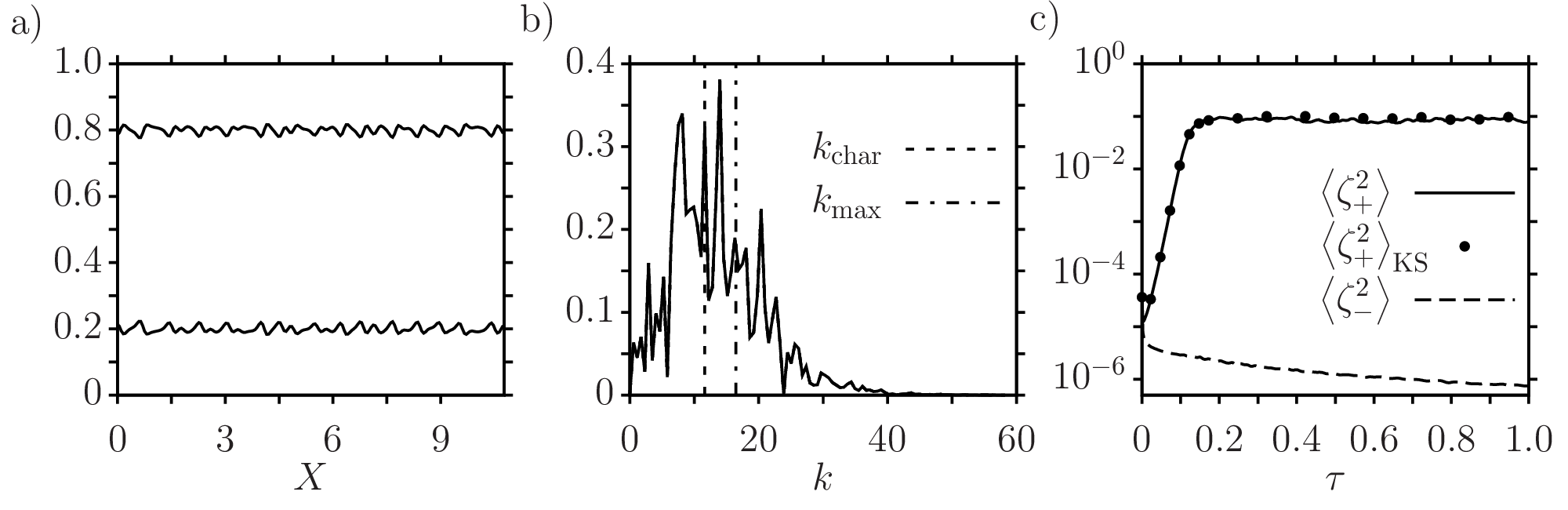}}\caption{An example for the numerical results obtained by solving \eqref{eq:ev_eg_sym_1} and \eqref{eq:ev_eg_sym_2} for $H_0 = 0.2$, $\Rey = 23.46$, $\Ca = 0.15$. Plot a) shows the interfaces along the channel at $\tau=1$, b) depicts the absolute values of the Fourier coefficients of $\zeta_+$ at $\tau=1$, and c) shows the time evolution of the symmetrical and the antisymmetrical instability mode. The approximate solution according to the Kuramoto-Sivashinsky (KS) equation is shown for comparison.}\label{fig:example}
\end{figure}

An example of the evolution of $\left<\zeta_-^2\right>$ and $\left<\zeta_+^2\right>$ is depicted in figure \ref{fig:example}. As for all of the simulations conducted, the length of the simulated domain was $L=40\pi/k_{\text{char}}$ with periodic boundary conditions. This corresponds to a sampling rate of $k_{\text{char}}/20$ in Fourier space. The length was divided into $200$ cells. Hence, according to the Nyquist theorem, the maximal numerically resolvable wavenumber would be $5k_{\text{char}}$ if the domain was discretised with a finite-different scheme. Nevertheless, the finite-element discretisation with quadratic Lagrangian shape functions utilized for the simulations exceeds this resolution. In any case, although the strong non-linearity of the evolution equations suggests that higher wavenumbers than $k_\text{max}=\sqrt{2}k_{\text{char}}$ should also appear during the evolution, the highest unstable wavenumber in the full non-linear simulations was still found to be considerably smaller than $5k_{\text{char}}$. As an initial condition, the interfaces were assumed to be flat except for a small white-noise perturbation with an amplitude of $0.01H_0$ added to the surface of the lower film. Since the upper film was initially undeformed ($\eta_2=0$), it was ensured that at the beginning $\left<\zeta_-^2\right>=\left<\zeta_+^2\right>$.

For the simulations referred to in figure \ref{fig:example}, silicone oil ($10\text{\,}\text{cSt}$) and air at a temperature of $50\text{\,}^\circ\text{C}$ and at $1\text{\,bar}$ pressure were assumed as the working fluids. Their material properties are summarized in table \ref{tab:mat_prop}. The undisturbed film thicknesses were $H_0=0.2$, which corresponds to an error of $\Delta\sigma_+\approx0.005$ and $\Delta\sigma_-\approx0.006$ in the (linear) growthrates caused by the truncation of the viscosity-ratio expansion (see equation \ref{eq:err_visc}). Therefore, the assumption of  semi-rigid liquid films is valid for this configuration. The channel width was set to $d=200\text{\,}\mu \text{m}$, and the mean air-flow velocity was assumed to be $3.5\text{\,m\,s}^{-1}$. The corresponding Reynolds and Capillary numbers are $\Rey\approx23.46$ and $\Ca \approx0.15$, while $\epsilon\approx 0.12$. The maximal deformation of the interfaces was found to be $\Delta H\approx0.028$; thus - according to equation \eqref{eq:eff_inflow} - the relative contribution to the pressure field due to inertia was approximately $0.23$. Hence, all assumptions necessary for the validity of equations \eqref{eq:ev_eg_sym_1} and \eqref{eq:ev_eg_sym_2} were fulfilled during the simulation.

Plot a) of figure \ref{fig:example} shows the interface shapes at $\tau = 1$, while plot b) depicts the spectrum of the absolute values of the Fourier coefficients of the symmetrical deformation, $\zeta_+$, at the same time instance. From this plot one can deduce that there is no well-defined characteristic (i.e. dominating) wavelength of the instability, which is a consequence of the pronounced non-linearity of the evolution equations. Similarly, while high wavenumbers are damped, the non-linearity of the full equations \eqref{eq:ev_eg_sym_1} and \eqref{eq:ev_eg_sym_2} implies that wavenumbers larger than $k_\text{max}$ (obtained from the linear analysis) are unstable as well. This does not cause any problems in the simulations, as the largest appearing wavenumber is still much smaller than the resolution of the mesh, $5k_{\text{char}}$.

Figure \ref{fig:example} c) shows the time evolution of $\left<\zeta_+^2\right>$ and $\left<\zeta_-^2\right>$. As one expects, the antisymmetrical instability mode is damped quickly, and $\left<\zeta_+^2\right>$ becomes much larger than $\left<\zeta_-^2\right>$ after a short initial transition period. For comparison, a simulation of the Kuramoto-Sivashinsky equation \eqref{eq:zeta_ks} was also performed. As initial condition, the initial distribution of $\zeta_+$ from the previous simulation was utilized (i.e. the initial white-noise distribution is identical). One finds that the graph of $\left<\zeta_+^2\right>$ obtained from the full simulation is essentially identical to the one calculated from the simplified problem $\left<\zeta_+^2\right>_{\text{KS}}$. This comparison was also performed for three other systems. For the first one, all parameters were the same as in figure \ref{fig:example}, but the mean air velocity was changed to $u=5\text{\,}\text{m} \text{\,s}^{\text{-}1}$. Similarly, for the second and third simulation, the governing parameters were unchanged, except that the non-dimensional film thickness was $H_0=0.1$ and the channel width was $d=100\text{\,}\mu\text{m}$, respectively. For all cases, the simplified equation \eqref{eq:zeta_ks} was found to provide a good approximation to the behaviour of the full system, even at early times. The accuracy of the approximation is comparable to what is shown in figure \ref{fig:example} c).

Returning to the question posed at the beginning of section \ref{sec:sym_sys}, one may conclude that the mirror symmetry of the system does not break down dynamically. Even if the fluctuations along the interfaces are not symmetrical, they do not give rise to a self-sustaining asymmetrical deformation of the interfaces. However, it should be noted that if the viscosity ratio $M$ is not sufficiently large, then this conclusion does no longer hold, since both the symmetrical and the antisymmetrical mode can become unstable. This is apparent from the formulas given in appendix \ref{app_growth}.

\section{Asymmetrical systems}\label{sec:asym_sys}

In realistic systems, complete mirror symmetry of the undisturbed system is not a valid assumption. For such cases it is possible to perform a second perturbation analysis for the deviation from the symmetrical configuration. A similar problem was discussed by \cite{papaef_2013}, and a similar solution approach could be applied herein. Nevertheless, in order to get a qualitative overview of systems with high asymmetry, it is favourable to use numerical calculations. These are discussed in the following.

Since the flow-pattern map of the asymmetrical system is expected to be highly complex and the equations are dependent on multiple independent parameters, it is not possible to probe the whole parameter space of the system with numerical simulations. Thus, the numerical analysis was restricted to setups, where the bottom and the top liquid films have the same material properties. Nonetheless, their initial thicknesses were allowed to differ by a non-dimensional quantity $\delta$, which defines the difference between the initial thicknesses of the undisturbed films according to
\vspace{-2pt}
\begin{equation}
H_1(t=0)=H_0(1+\delta/2), \quad \quad \quad H_2(t=0)=H_0(1-\delta/2) \quad \quad \quad \delta \in[0,2).
\end{equation}

\noindent The goal of the present analysis is to find how the coupling between the two films changes with $\delta$. For this purpose, the evolution of the liquid films was simulated with the same numerical method as introduced in the previous section. Furthermore, the quantities used for non-dimensionalisation were defined in the same way as for symmetrical systems. The advantage of using identical liquid material properties is that the evolution of the film thicknesses is still described by equations \eqref{eq:ev_eg_sym_1} and \eqref{eq:ev_eg_sym_2}. Therefore, the dynamics of the interfaces should be only governed by the four dimensionless numbers, $\epsilon$, $\Rey\sqrt[3]{Ca}$, $H_0$ and $\delta$.

To quantify the coupling strength between the layers, the normalized correlation of the interface deformation amplitudes $\eta_1$ and $\eta_2$ is defined by expression \eqref{eq:deff_corr}.

\newpage

\begin{equation}\label{eq:deff_corr}
C(\tau)=\frac{\int\limits_{-\infty}^{\infty}{\eta_1(\tau,X)\eta_2(\tau,X)}dX}{\sqrt{\int\limits_{-\infty}^{\infty}{\eta_1^2(\tau,X)dX}\int\limits_{-\infty}^{\infty}{\eta_2^2(\tau,X)dX}}}.
\end{equation}

\noindent A convenient property of $C$ is that it is invariant under the transformation $\eta_1\rightarrow\alpha_1\eta_1$, $\eta_2\rightarrow\alpha_2\eta_2$ as long as $\alpha_1$ and $\alpha_2$ are independent of  $X$. Thus, even if the magnitudes of the deformations are different, the normalized correlation function may remain the same. This is useful since one expects the thinner film to have a smaller deformation than the thicker one.

All simulations indicate that at $\delta=0$ the correlation function tends to unity as $\tau \rightarrow \infty$. For symmetrical systems this is to be expected, since - as discussed above - in those cases the antisymmetrical mode disappears in the long-time limit. Hence, for symmetrical systems $\eta_1\approx\eta_2$ after a sufficiently long time period. Relative to the symmetrical case, the correlation is seen to decrease with increasing values of $\delta$. However, one finds that $C$ stays positive for all values of $\delta$. A number of simulations were performed for the non-dimensional parameters $H_0\in[0.05,0.2]$, $Re\sqrt[3]{Ca}\in[1,10]$, $\delta \in[0,0.95]$ and $\epsilon \in [0.01,0.125]$. The value of $C$ was found to be non-negative in all cases studied. Consequently, even for the asymmetrical configuration, the two films prefer to deform in phase with each other.

\begin{figure}
\centerline{\includegraphics[width=0.5\linewidth]{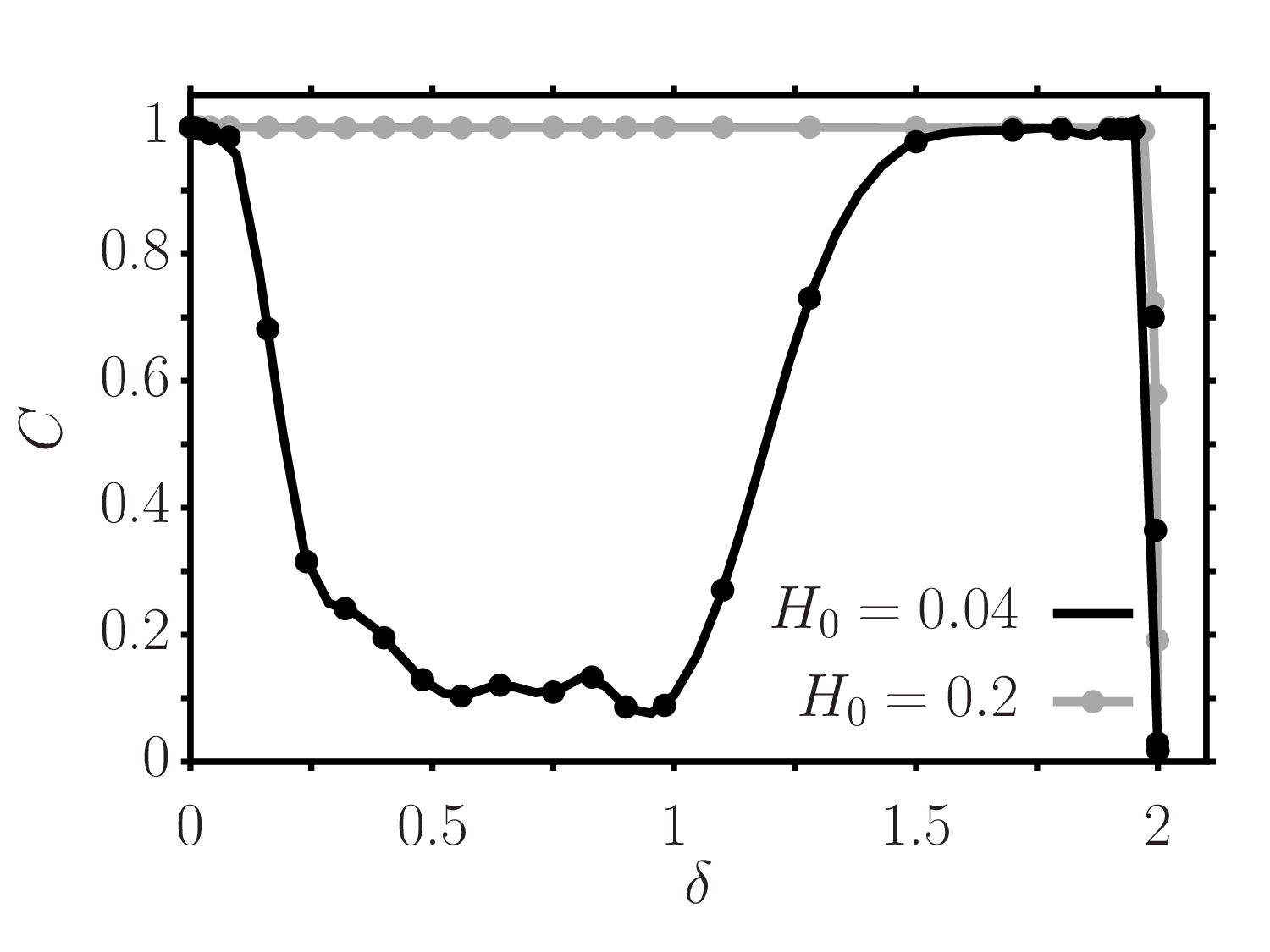}}
\caption{The dependence of the correlation function $C$ on the thickness difference $\delta$ for systems with different nominal film thicknesses $H_0$.}\label{fig:numeric}
\end{figure}

Two examples of the evolution of the correlation function with $\delta$ are shown in figure \ref{fig:numeric}. Once again the silicone oil of table \ref{tab:mat_prop} was considered as the liquid medium, while air was chosen for the intermediate gas layer. In the first simulation the channel thickness was $d=50\text{\,}\mu \text{m}$. The dimensionless initial liquid layer thicknesses for $\delta=0$ were $H_0=0.04$, and the mean air velocity was $u=5\text{\,m\,s}^{\text{-}1}$. In the second one, these parameters were $d=100\text{\,}\mu\text{m}$, $H_0=0.2$ and $u=3\text{\,m\,s}^{\text{-}1}$. In dimensionless coordinates, the channel length was set equal to $20$, while it was divided into at least 500 cells to ensure grid independence for both simulations. The simulated time interval after which $C$ was evaluated was chosen to be long enough to ensure that the values of $\left<\eta_1^2\right>$ and $\left<\eta_2^2\right>$ only oscillate around an asymptotic value. The corresponding time was $1600$ in the first and $8$ in the second simulation. This corresponds to $849\text{\,s}$ and $606\text{\,s}$ in real time. The correlation function was calculated as the average of the correlation between the film thicknesses in the final $10\,\%$ of the simulated time interval.

For $H_0=0.2$ it is found that the correlation of the two films remains close to unity even for relatively high values of $\delta$. However at $\delta \approx 2$, when the ratio of the undisturbed film thicknesses is about $100$, $C$ drops down sharply. This is not surprising, since at this point the order of magnitude of the terms in the evolution equations \eqref{eq:ev_eg_sym_1} and \eqref{eq:ev_eg_sym_2} are principally different for each film, and one does no longer expect the layers to have a qualitatively similar behaviour. Nevertheless, it is remarkable that up to this point the coupling between the films is sufficiently strong to force the film deformations to be in phase with each other. 

An example for a less common dynamical behaviour was found for $H_0=0.04$. Here it is observed that the correlation drops sharply already at smaller values of $\delta$. This was found to be a consequence of a secondary instability, possibly the Majda-Pego instability, growing at a slower rate and mainly affecting the pattern of the thinner layer. However, as $\delta$ further increases this instability either disappears or its growthrate becomes small enough to no longer affect the patterns for the simulated time interval, leading to an increase in the correlation between the films at about $\delta \approx 2$.

All other numerical simulations showed a qualitatively similar behaviour as either of the two cases shown in figure \ref{fig:numeric}. The correlations of most systems were close to unity for $\delta \in [0,2)$, with a small drop in its value at intermediate $\delta$. However, no other system has been found for which the correlation approaches $0$ so closely as for $H_0=0.04$ in figure \ref{fig:numeric}. Surprisingly, this indicates that even in a highly asymmetrical system, one may often utilize the assumption that the patterns on the two interfaces are identical apart from a constant factor between the amplitudes. This high level of phase synchronization is especially remarkable since the power spectrum of the deformations does not indicate phase coherence \citep{boccaletti_synchronized}, which can be exemplarily deduced from figure \ref{sec:sym_sys}b. For highly asymmetrical systems the deformation of the thinner film is expected to have a smaller effect on the pattern of the thicker film than vice versa. Therefore, at high values of $\delta$ where $C\approx 1$ for both cases shown in figure \ref{fig:numeric}, this indicates that the deformation of the thinner film is enslaved by the instability of the thicker one.

\section{Conclusion}

In this paper the long-wavelength instabilities emerging at the interfaces of thin liquid films, which cover the walls of a plane-parallel channel and are exposed to a gaseous flow along the channel centre, has been investigated. The corresponding equations were simplified by employing the long-wavelength approximation and by assuming moderate values for the Reynolds number. The small viscosity ratios of the gaseous to the liquid mediums were utilized to simplify the system further, so that it can be described and analysed by two relatively compact evolution equations. 

Based on this simplified model it was shown that for symmetrical films on both walls (i.e. identical material properties and initial thicknesses) the interface deformations can be described by the superposition of a symmetrical and a linearly independent antisymmetrical mode. By means of a linear and weakly non-linear stability analysis it was found that the symmetrical mode is linearly unstable but grows only up  to a maximum deflection amplitude. Given the strong non-linearity of the original set of film evolution equations, the linearly fastest growing wavenumber and the linearly largest unstable wavenumber were demonstrated to be only rough orientation marks for the wavenumber spectrum obtained numerically. The linear analysis showed that for liquid films thinner than $1/8$ of the total channel width, the instability is mainly driven by variations of the viscous stresses along the interfaces. For thicker films variations in the pressure gradient are the main source of the instability. By contrast to the symmetrical mode, the antisymmetrical mode is not inherently unstable but can only exhibit an instability due to its coupling to the symmetrical mode. Nevertheless, at long times the antisymmetrical mode disappears, so that in the weakly non-linear regime the system can be described by a single Kuramoto-Sivashinsky equation. However, this observation does not necessarily hold for arbitrary viscosity ratios. For the latter case, analytical expressions for the (linear) growthrates are provided, which were found to be in complete agreement with numerical simulations conducted for selected cases by other researchers. Furthermore, numerical simulations were performed for asymmetrical systems. For long elapsed time periods it was found that typically the correlation between the interfacial deformations remains close to unity even for highly asymmetrical configurations. This implies that even for such systems the patterns on the two interfaces remain approximately phase-synchronized. Nevertheless, for certain values of the characteristic parameters, a secondary instability can slowly emerge, disrupting the highly correlated patterns of the two films.

Finally, it should be mentioned that the method used for the analysis of this system can be easily generalized to more complex systems. For instance, introducing Marangoni stresses or gravitational forces to the equations will only add a new term to the evolution equations \eqref{eq:ev_eg_sym_1} and \eqref{eq:ev_eg_sym_2}, but the approach to tackle these problems is essentially the same as the one discussed in this paper. Our findings are relevant for systems where the walls of narrow ducts are coated by thin liquid films, which, however, do not severely obstruct the gas flow in the centre. Examples are air flow in respiratory ducts coated with mucus, boiling in microchannels, or the flow of a supersaturated gas in a narrow duct such that liquid condenses at the channel walls. 

This study was supported by the Deutsche Forschungsgemeinschaft (DFG) under grant number DI 1689/1-1, which is gratefully acknowledged.

\appendix
\section{}\label{app_growth}
The growthrates from the linear stability analysis of section \ref{sec:lin_stab_anal} for arbitrary viscosities are given by

\begin{align}
\text{Re}(\sigma_{+})=&\frac{\left(1-2 H_0\right){}^2 H_0^2}{210 \left(\Mrat-6 (\Mrat-1) H_0+12 (\Mrat-1) H_0^2-8 (\Mrat-1) H_0^3\right){}^5}\nonumber\\
&\times{\Big\{}k^2 18 \text{Ca}^{1/3} (\Mrat-1) \Rey \left(1-2 H_0\right){\Big[}2 \Mrat^4-4 \Mrat^3 (-3+7 \Mrat) H_0\nonumber\\
&-\Mrat^2 (-63+4 (37-36 \Mrat) \Mrat) H_0^2-6 \Mrat^2 (105+4 \Mrat (-17+8 \Mrat)\nonumber\\
&-56\Rrat) H_0^3-2 \Mrat (2 \Mrat (-133+12 \Mrat (-51+28 \Mrat))+7 (-63+178 \Mrat)\Rrat) H_0^4\nonumber\\
&-(-756\Rrat+4 \Mrat (21 (60+49\Rrat)-2 \Mrat (1981+4 \Mrat (-673+252 \Mrat)+753\Rrat))) H_0^5\nonumber\\
&-12 (21 (15+7\Rrat)+\Mrat (-3255-179\Rrat+4 \Mrat (1743+4 \Mrat (-373+112 \Mrat)\nonumber\\
&+44\Rrat))) H_0^6-(48 (-420+53\Rrat)+16 \Mrat (7210-1027\Rrat+2 \Mrat (-5999\nonumber\\
&+132 (31-8 \Mrat) \Mrat+422\Rrat))) H_0^7-(64 (-1+\Mrat) \Mrat (1925+28 \Mrat (-59+18 \Mrat)\nonumber\\
&-345\Rrat)+96 (-1+\Mrat) (-385+97\Rrat)) H_0^8-128 (-1+\Mrat) (210-54\Rrat\nonumber\\
&+\Mrat (-595+68 (7-2 \Mrat) \Mrat+99\Rrat)) H_0^9-64 (-1+\Mrat) (8 \Mrat (35-28 \Mrat\nonumber\\
&+8 \Mrat^2-5\Rrat)+5 (-21+5\Rrat)) H_0^{10}{\Big]}+k^4 35 H_0 {\Big[}(4\Mrat-3) H_0-2\Mrat{\Big]} {\Big[}\Mrat\nonumber\\
&-2(\Mrat-1) H_0 \left(3-6 H_0+4 H_0^2\right){\Big]}{}^4{\Big\}},
\end{align}\vspace{-6pt}
\begin{align}
\text{Re}(\sigma_{-})=&\frac{H_0^3}{30 \text{Ca}^{2/3} \left(\Mrat-2 (-1+\Mrat) H_0\right){}^3 \left(\Mrat-2 (-1+\Mrat) H_0 \left(3-6 H_0+4 H_0^2\right)\right){}^2}\nonumber\\
&\times{\Big\{}k^2{\Big[}18 \text{Ca} (-1+\Mrat) \Rey \left(1-2 H_0\right) H_0 {\Big(}5 \Mrat^2+2 \Mrat^2 (-15+8\Rrat) H_0\nonumber\\
&+2 \Mrat (\Mrat (30-32\Rrat)+7\Rrat) H_0^2+4 (R-7 \Mrat\Rrat+2 \Mrat^2 (-5+8\Rrat)) H_0^3{\Big)}{\Big]}\nonumber\\
&+k^4{\Big[}5 \text{Ca}^{2/3} \left(-2 \Mrat+(-1+4 \Mrat) H_0\right) {\Big(}\Mrat^2-8 (-1+\Mrat) \Mrat H_0+12 (1-3 \Mrat\nonumber\\
&+2 \Mrat^2) H_0^2-8 \left(3-7 \Mrat+4 \Mrat^2\right) H_0^3+16 (-1+\Mrat)^2 H_0^4{\Big)}{}^2{\Big]}{\Big\}},
\label{eq:general_visc2}
\end{align}

\noindent where $R$ is the ratio of the mass densities, defined in section \ref{sec:ev_eq}.

These formulas were compared with the results of the numerical calculations of \cite{renardy_triple_layer}. The corresponding results are summarized in figure \ref{fig:renardy}. In order to match with the non-dimensional parameters used in the paper referred to, the material properties were taken as $\mu_1=\mu_3=1.56605\times10^{\text{-}6}\text{ Pa\,s}$, $\rho_1=\rho_2=\rho_3=1\text{ kg\,m}^{\text{-}3}$, and a channel width of $200\text{ }\mu \text{m}$ was chosen. \cite{renardy_triple_layer} has performed her numerical calculations for vanishing surface tension. However, the scaling introduced in the current paper is based on a finite capillary number. For this purpose the surface tension of the silicone oil listed in table \ref{tab:mat_prop} was used. Nevertheless, the effects of the surface tension can be eliminated subsequently by omitting the terms scaling with $k^4$ in the growthrates \eqref{eq:general_visc1} and \eqref{eq:general_visc2}. In figure \ref{fig:renardy}, the growthrates for the wavenumber $0.01/H_0$ are plotted for a pressure gradient of $\partial p/\partial x=19.62\text{\,Pa\,m}^{\text{-}1}$ as a function of the non-dimensional film thickness.

\cite{papaef_2013} discussed that the weakly non-linear equations describing systems of arbitrary viscosity ratios should still take a qualitatively similar form to equations \eqref{eq_weak_nonlin1} and
\eqref{eq_weak_nonlin2}. In fact, expression \eqref{eq:general_visc2} indicates that for arbitrary viscosities the antisymmetrical mode is not necessarily linearly stable. Therefore, for such systems the mirror symmetry can break down dynamically. A similar feature for the related problem of core-annular flows was obtained numerically by \cite{hu_asym_core_ann}.

\section{}\label{ampl_eq}

Most linear terms of equations \eqref{eq_weak_nonlin1} and \eqref{eq_weak_nonlin2} drop out after substitution of the ansatz for the solution \eqref{eq:general_sol}. This is a direct consequence of employing a generalized form of the solution used for the linearized equation to describe the weakly non-linear behaviour. For instance, equation \eqref{eq_weak_nonlin1} takes the form

\begin{align}\label{eq:ew_amp_weakly_nonlin}
&\int\limits_{-\infty}^{\infty}{{\Bigg(}\frac{\partial A_+}{\partial \tau}e^{\sigma_+\tau}+\frac{\partial A_-}{\partial \tau}e^{\sigma_-\tau}{\Bigg)}e^{\iu kX}dk}\nonumber\\
&\quad\quad=-\frac{\partial}{\partial X}{\Bigg[}\left\{\int\limits_{-\infty}^{\infty}{\left(A_+e^{\sigma_+\tau}+A_-e^{\sigma_-\tau}\right)e^{\iu kX}dk}\right\}^2\frac{3(1-H_0)[1+H_0(3-2H_0)]}{(1-2H_0)^5}\nonumber\\
&\quad\quad+\left\{\int\limits_{-\infty}^{\infty}{\left(A_+e^{\sigma_+\tau}+A_-e^{\sigma_-\tau}\right)e^{\iu kX}dk}\right\}\left\{\int\limits_{-\infty}^{\infty}{\left(A_+e^{\sigma_+\tau}-A_-e^{\sigma_-\tau}\right)e^{\iu k'X}dk'}\right\}\nonumber\\
&\quad\quad\times\frac{6H_0(2+H_0-2H_0^2)}{(1-2H_0)^5}+\left\{\int\limits_{-\infty}^{\infty}{\left(A_+e^{\sigma_+\tau}-A_-e^{\sigma_-\tau}\right)e^{\iu kX}dk}\right\}^2\frac{3H_0^2(3+2H_0)}{(1-2H_0)^5}{\Bigg]}.
\end{align}

Since multiplication in the position space transforms to convolution in Fourier space, one finds that

\begin{equation}\label{eq:ew_amp_weakly_nonlin2}
\begin{aligned}&\frac{\partial A_+}{\partial \tau}e^{\sigma_+\tau}+\frac{\partial A_-}{\partial \tau}e^{\sigma_-\tau}\\
&\quad\quad=-\iu k{\Bigg[}\frac{3(1+6H_0)}{(1-2H_0)^5}\left(A_+e^{\sigma_+\tau}\right)\underset{k}*\left(A_+e^{\sigma_+\tau}\right)+\frac{6(1+4H_0)}{(1-2H_0)^4}\left(A_+e^{\sigma_+\tau}\right)\underset{k}*\left(A_-e^{\sigma_-\tau}\right)\\
&\quad\quad+\frac{3(1+2H_0)}{(1-2H_0)^3}\left(A_-e^{\sigma_-\tau}\right)\underset{k}*\left(A_-e^{\sigma_-\tau}\right){\Bigg]}.
\end{aligned}
\end{equation}

\noindent A similar calculation can be performed for equation \eqref{eq_weak_nonlin2}. A straightforward simplification of these two equations leads to the form of equations \eqref{eq:amp_ev}.

\section{}\label{kin_inst}

There are two known instabilities appearing during stratified film flows at zero Reynolds numbers. The first one is the kinetic alpha effect which appears from the resonance-like coupling of the interfaces. According to the results of \cite{kliakhandler_alpha}, this instability is significant for configurations where the intermediate layer is thin. However, assumption 3 of section \ref{sec:framework} is only justified if the mean gas velocity is much larger than the velocities at the gas-liquid interfaces. This is not fulfilled for thin gas layers. Thus, the kinetic alpha effect is automatically excluded from the analysis described in this paper.

Nevertheless, a different kinetic instability could appear in the system, as was also indicated by \cite{kliakhandler_alpha} as well as by \cite{papaef_2013}. This instability is induced by capillary forces, and it is suspected to be a fourth-order generalization of the Majda-Pego instability \citep{majda-pego_instability} associated with second-order dissipative systems. As it was mentioned in the discussion of equations \eqref{eq:nonlin_av1} and \eqref{eq:nonlin_av2}, in symmetrical systems the surface tension always decreases the deformations of the interfaces, thus no Majda-Pego instability appears. This agrees with the expectations, as the instability criteria discussed by \cite{canic_majda-pego} and assumed by \cite{papaef_2013} to be valid for quadratic dissipative systems cannot be fulfilled. Formally, according to equations \eqref{eq_weak_nonlin1} and \eqref{eq_weak_nonlin2}, the dissipation matrix of the Majda-Pego instability is defined 
by

\begin{equation}
\mathsfbi{D}=\begin{pmatrix} H_0^3/3 & 0\\ 0& H_0^3/3
\end{pmatrix}.
\end{equation}

\noindent Since this is a positive multiple of the identity matrix, the system cannot be unstable in the sense of Majda and Pego. However, for asymmetrical systems the diagonal elements of $\mathsfbi{D}$ are no longer equal, thus the kinetic instability may appear. We suspect that this is the source of the secondary instability observed in figure \ref{fig:numeric}.

\newpage

\section{}\label{reduction}
Equations \eqref{eq:wn_zetap} and \eqref{eq:wn_zetam} are reducible to a single evolution equation, even without assuming that $\zeta_-\ll\zeta_+$. In order to show this one introduces 
\begin{equation}
Z_-(Y)=\int\limits_{-\infty}^Y{\zeta_-}dX.
\end{equation}

\noindent After integrating equation \eqref{eq:wn_zetam} with respect to $X$, one finds

\begin{equation}
\zeta_+\Big|_X=-\left\{\frac{(1-2H_0)^4}{6(1+4H_0)Z_-}\left[\frac{1}{\epsilon}\frac{\partial Z_-}{\partial \tau}+\frac{H_0^3}{3}\frac{\partial^4Z_-}{\partial X^4}\right]\right\}{\Bigg |}_{X+\frac{12H_0^2}{(1-2H_0)^4}\tau}.
\end{equation}

\noindent Substituting this into \eqref{eq:wn_zetap} and using $\zeta_-=\partial Z_-/\partial X$ one arrives at a single evolution equation for $Z_-$. The resulting equation is considerably more complicated than the Kuramoto-Sivashinsky equation. However, after solving this single equation one can directly calculate the deformations of the interfaces. 

As an equivalent approach, one can also reduce the number of equations by integrating equation \eqref{eq:wn_zetap} instead of \eqref{eq:wn_zetam}.  Subsequently, $\zeta_-$ can be computed. This can be applied to arrive at a single evolution equation for the spatial integral of $\zeta_+$. Since generally $\zeta_-\ll\zeta_+$, in certain cases it may be beneficial to use this equation for the simulations in order to avoid numerical inaccuracies.

\section{}\label{simres}

Numerical simulations were performed in order to support the argument that the antisymmetrical mode disappears if the system is initially mirror symmetric. The parameters of the simulations are summarized in section \ref{sec:weak_nonlin}. The results are given in table \ref{tab:num_res}. The simulated time intervals were set to $\tau_{\text{max}}$, the value of which was chosen large enough for $\left<\zeta_+^2\right>$ to reach an asymptotic state, i.e. it just slightly oscillates around its asymptotic value. At the last time step the ratio of $\left<\zeta_+^2\right>$ and $\left<\zeta_-^2\right>$ is calculated. The results confirm that after a sufficiently long time the magnitude of the antisymmetrical mode is considerably smaller than that of the symmetrical one.

\begin{table}
  \centering
    \begin{tabular}{p{0.75cm}  p{0.75cm}  p{0.75cm}  p{0.75cm} p{0.75cm} p{1.5cm} p{0.06cm}|}
    $H_0$		&	$\Rey$ 	&	$\Ca$		&	$\epsilon$	& $\tau_{\text{max}}$	&	$\left<\zeta_+^2\right>/\left<\zeta_-^2\right>$ &\\
    $0.01$	&	$10$		&	$0.5$		&	$0.1$				&	$1200$							&	$6.76\times10^2$ &\\
    $0.02$	&	$10$		&	$0.5$		&	$0.1$				&	$800$								&	$1.51\times10^3$ &\\
    $0.05$	&	$10$		&	$0.5$		&	$0.1$				&	$400$								&	$6.02\times10^3$ &\\
    $0.1$		&	$10$		&	$0.5$		&	$0.1$				&	$40$								&	$1.13\times10^4$ &\\
    $0.2$		&	$10$		&	$0.5$		&	$0.1$				&	$20$								&	$3.51\times10^4$ &\\
    $0.3$		&	$10$		&	$0.5$		&	$0.1$				&	$2$									&	$3.22\times10^4$ &\\
    $0.1$		&	$1$			&	$0.5$		&	$0.1$				&	$500$								&	$1.56\times10^1$ &\\
    $0.1$		&	$2$			&	$0.5$		&	$0.1$				&	$80$								&	$8.25\times10^4$ &\\
    $0.1$		&	$5$			&	$0.5$		&	$0.1$				&	$80$								&	$8.92\times10^4$ &\\
    $0.1$		&	$10$		&	$0.5$		&	$0.005$			&	$20$								&	$1.89\times10^1$ &\\
    $0.1$		&	$20$		&	$0.5$		&	$0.1$				&	$80$								&	$6.70\times10^5$ &\\
    $0.1$		&	$50$		&	$0.5$		&	$0.1$				&	$60$								&	$8.03\times10^7$ &
    \end{tabular}~~
    \begin{tabular}{p{0.75cm}  p{0.75cm}  p{0.75cm}  p{0.75cm} p{0.75cm} p{1.5cm}}
    $H_0$		&	$\Rey$ 	&	$\Ca$		&	$\epsilon$	& $\tau_{\text{max}}$	&	$\left<\zeta_+^2\right>/\left<\zeta_-^2\right>$\\
    $0.1$		&	$10$		&	$0.125$	&	$0.1$				&	$20$								&	$1.41\times10^3$\\
    $0.1$		&	$10$		&	$0.25$	&	$0.1$				&	$20$								&	$7.09\times10^3$\\
    $0.1$		&	$10$		&	$1$			&	$0.1$				&	$20$								&	$1.59\times10^4$\\
    $0.1$		&	$10$		&	$2$			&	$0.1$				&	$20$								&	$4.43\times10^4$\\
    $0.1$		&	$10$		&	$5$			&	$0.1$				&	$40$								&	$6.93\times10^4$\\
    $0.1$		&	$10$		&	$0.5$		&	$0.005$			&	$40$								&	$1.94\times10^1$\\
    $0.1$		&	$10$		&	$0.5$		&	$0.01$			&	$40$								&	$8.35\times10^1$\\
    $0.1$		&	$10$		&	$0.5$		&	$0.02$			&	$40$								&	$2.90\times10^2$\\
    $0.1$		&	$10$		&	$0.5$		&	$0.05$			&	$40$								&	$2.12\times10^3$\\
    $0.1$		&	$10$		&	$0.5$		&	$0.2$				&	$40$								&	$6.29\times10^4$\\
    $0.1$		&	$10$		&	$0.5$		&	$0.5$				&	$40$								&	$6.65\times10^5$\\
    &
    \end{tabular}
    \caption{Results of the numerical simulations for symmetrical configurations. The last column contains the ratio of the magnitudes of the symmetrical and the antisymmetrical modes at $\tau=\tau_{\text{max}}$.}
	  \label{tab:num_res}
\end{table}


\begin{thebibliography}{38}
\expandafter\ifx\csname natexlab\endcsname\relax\def\natexlab#1{#1}\fi
\def\au#1{#1} \def\ed#1{#1} \def\yr#1{#1}\def\at#1{#1}\def\jt#1{\textit{#1}}
  \def\bt#1{#1}\def\bvol#1{\textbf{#1}} \def\vol#1{#1} \def\pg#1{#1}
  \def\publ#1{#1}\def\arxiv#1{#1}\def\org#1{#1}\def\st#1{\textit{#1}}

\bibitem[Anturkar {\em et~al.\/}(1990)Anturkar, Papanastasiou \&
  Wilkes]{aturkar_lin_multi}
{\sc \au{Anturkar, N.~R.}, \au{Papanastasiou, T.~C.} \& \au{Wilkes, J.~O.}}
  \yr{1990}  \at{Linear stability analysis of multilayer plane {P}oiseuille
  flow}.  \jt{Physics of Fluids A: Fluid Dynamics (1989-1993)}  \bvol{2}~(4),
  \pg{530--541}.

\bibitem[Boccaletti(2008)]{boccaletti_synchronized}
{\sc \au{Boccaletti, S}} \yr{2008} {\em The synchronized dynamics of complex
  systems\/}.  \publ{Elsevier}.

\bibitem[Canic \& Plohr(1995)]{canic_majda-pego}
{\sc \au{Canic, S.} \& \au{Plohr, B.J.}} \yr{1995}  \at{Shock wave
  admissibility for quadratic conservation laws}.  \jt{Journal of Differential
  Equations}  \bvol{118}~(2),  \pg{293 -- 335}.

\bibitem[Chandrasekhar(2013)]{chandrasekhar_hydrodynamic}
{\sc \au{Chandrasekhar, Subrahmanyan}} \yr{2013} {\em Hydrodynamic and
  hydromagnetic stability\/}.  \publ{Courier Corporation}.

\bibitem[Chang \& Demekhin(2002)]{chang_complex_wave}
{\sc \au{Chang, Hen-hong} \& \au{Demekhin, Evgeny~A}} \yr{2002} {\em Complex
  wave dynamics on thin films\/}.  \publ{Elsevier}.

\bibitem[Charru \& Hinch(2006)]{charru_sand_ripple}
{\sc \au{Charru, F.} \& \au{Hinch, E.~J.}} \yr{2006}  \at{Ripple formation on a
  particle bed sheared by a viscous liquid. part 1. steady flow}.  \jt{Journal
  of Fluid Mechanics}  \bvol{550},  \pg{111--121}.

\bibitem[Comsol(2014)]{comsol}
{\sc \au{Comsol}} \yr{2014} COMSOL
  Multiphysics$\textsuperscript{\textregistered}$, COMSOL, Inc., G\"ottingen,
  Germany.

\bibitem[{Cvitanovic} {\em et~al.\/}(2010){Cvitanovic}, {Davidchack} \&
  {Siminos}]{cvitanovic_phase_space}
{\sc \au{{Cvitanovic}, P.}, \au{{Davidchack}, R.~L.} \& \au{{Siminos}, E.}}
  \yr{2010}  \at{{On the State Space Geometry of the {K}uramoto--{S}ivashinsky
  Flow in a Periodic Domain}}.  \jt{SIAM Journal on Applied Dynamical Systems}
  \bvol{9},  \pg{1--33}.

\bibitem[Drazin \& Reid(2004)]{drazin_hydrodynamic}
{\sc \au{Drazin, P.~G.} \& \au{Reid, W.~H.}} \yr{2004} {\em Hydrodynamic
  stability\/}.  \publ{Cambridge University Press}.

\bibitem[Heil {\em et~al.\/}(2008)Heil, Hazel \& Smith]{heil_airway_closure}
{\sc \au{Heil, M.}, \au{Hazel, A.~L.} \& \au{Smith, J.~A.}} \yr{2008}  \at{The
  mechanics of airway closure}.  \jt{Respiratory Physiology \& Neurobiology}
  \bvol{163}~(1–3),  \pg{214 -- 221}, {R}espiratory {B}iomechanics.

\bibitem[Hewitt \& Hall-Taylor(1970)]{hewitt_annular_flow}
{\sc \au{Hewitt, G.F.} \& \au{Hall-Taylor, N.S.}} \yr{1970} {\em Annular
  Two-phase Flow\/}.  \publ{Pergamon}.

\bibitem[Hooper \& Grimshaw(1985)]{hooper_double_nonlin}
{\sc \au{Hooper, A.~P.} \& \au{Grimshaw, R.}} \yr{1985}  \at{Nonlinear
  instability at the interface between two viscous fluids}.  \jt{Physics of
  Fluids (1958-1988)}  \bvol{28}~(1),  \pg{37--45}.

\bibitem[Hu \& Patankar(1995)]{hu_asym_core_ann}
{\sc \au{Hu, H.~H.} \& \au{Patankar, N.}} \yr{1995}  \at{Non-axisymmetric
  instability of core-annular flow}.  \jt{Journal of Fluid Mechanics}
  \bvol{290},  \pg{213--224}.

\bibitem[Hyman \& Nicolaenko(1986)]{hyman_kuramoto_limit}
{\sc \au{Hyman, J.~M.} \& \au{Nicolaenko, B.}} \yr{1986}  \at{The
  {K}uramoto-{S}ivashinsky equation: A bridge between pde's and dynamical
  systems}.  \jt{Physica D: Nonlinear Phenomena}  \bvol{18}~(1–3),  \pg{113
  -- 126}.

\bibitem[Johnson {\em et~al.\/}(1991)Johnson, Kamm, Ho, Shapiro \&
  Pedley]{johnson_airway_collapse}
{\sc \au{Johnson, M.}, \au{Kamm, R.~D.}, \au{Ho, L.~W.}, \au{Shapiro, A.~H.} \&
  \au{Pedley, T.~J.}} \yr{1991}  \at{The nonlinear growth of
  surface-tension-driven instabilities of a thin annular film}.  \jt{Journal of
  Fluid Mechanics}  \bvol{233},  \pg{141--156}.

\bibitem[Kandlikar(2012)]{kandlikar_flow_boiling}
{\sc \au{Kandlikar, S.~G.}} \yr{2012}  \at{History, advances, and challenges in
  liquid flow and flow boiling heat transfer in microchannels: a critical
  review}.  \jt{Journal of Heat Transfer}  \bvol{134}~(3),  \pg{034001}.

\bibitem[Kevrekidis {\em et~al.\/}(1990)Kevrekidis, Nicolaenko \&
  Scovel]{kevrekedis_comp}
{\sc \au{Kevrekidis, I.~G.}, \au{Nicolaenko, B.} \& \au{Scovel, J.~C.}}
  \yr{1990}  \at{Back in the saddle again: A computer assisted study of the
  {K}uramoto-{S}ivashinsky equation}.  \jt{SIAM Journal on Applied Mathematics}
   \bvol{50}~(3),  \pg{pp. 760--790}.

\bibitem[Kliakhandler \& Sivashinsky(1995)]{kliakhandler_alpha}
{\sc \au{Kliakhandler, I.} \& \au{Sivashinsky, G.}} \yr{1995}  \at{Kinetic
  alpha effect in viscosity stratified creeping flows}.  \jt{Physics of Fluids}
   \bvol{7}~(8),  \pg{1866--1871}.

\bibitem[Kudryashov(1990)]{kudryashov_exact_sol}
{\sc \au{Kudryashov, N.A.}} \yr{1990}  \at{Exact solutions of the generalized
  {K}uramoto-{S}ivashinsky equation}.  \jt{Physics Letters A}
  \bvol{147}~(5–6),  \pg{287 -- 291}.

\bibitem[Kuramoto \& Tsuzuki(1976)]{kuramoto}
{\sc \au{Kuramoto, Y.} \& \au{Tsuzuki, T.}} \yr{1976}  \at{Persistent
  propagation of concentration waves in dissipative media far from thermal
  equilibrium}.  \jt{Progress of Theoretical Physics}  \bvol{55}~(2),
  \pg{356--369}.

\bibitem[Li(1969)]{li_three_layer}
{\sc \au{Li, C.~H.}} \yr{1969}  \at{{I}nstability of {T}hree--{L}ayer {V}iscous
  {S}tratified {F}low}.  \jt{Physics of Fluids (1958-1988)}  \bvol{12}~(12),
  \pg{2473--2481}.

\bibitem[Lide \& Haynes(2010)]{crc_handbook}
{\sc \au{Lide, D.~R.} \& \au{Haynes, W.~M.}}, ed. \yr{2010} {\em {{CRC}
  Handbook of {C}hemistry and {P}hysics}\/}, 90th edn.  \publ{CRC Press}.

\bibitem[Majda \& Pego(1985)]{majda-pego_instability}
{\sc \au{Majda, A.} \& \au{Pego, R.~L.}} \yr{1985}  \at{Stable viscosity
  matrices for systems of conservation laws}.  \jt{Journal of Differential
  Equations}  \bvol{56}~(2),  \pg{229 -- 262}.

\bibitem[Nicolis \& Prigogine(1977)]{nicolis_self_org}
{\sc \au{Nicolis, G.} \& \au{Prigogine, I.}} \yr{1977} {\em Self-Organization
  in Nonequilibrium Systems\/}.  \publ{Wiley}.

\bibitem[Oron {\em et~al.\/}(1997)Oron, Davis \& Bankoff]{oron_lubrication}
{\sc \au{Oron, A.}, \au{Davis, S.~H.} \& \au{Bankoff, S.~G.}} \yr{1997}
  \at{Long-scale evolution of thin liquid films}.  \jt{Rev. Mod. Phys.}
  \bvol{69},  \pg{931--980}.

\bibitem[Papaefthymiou {\em et~al.\/}(2013)Papaefthymiou, Papageorgiou \&
  Pavliotis]{papaef_2013}
{\sc \au{Papaefthymiou, E.~S.}, \au{Papageorgiou, D.~T.} \& \au{Pavliotis,
  G.~A.}} \yr{2013}  \at{Nonlinear interfacial dynamics in stratified
  multilayer channel flows}.  \jt{Journal of Fluid Mechanics}  \bvol{734},
  \pg{114--143}.

\bibitem[Reisfeld {\em et~al.\/}(1991)Reisfeld, Bankoff \&
  Davis]{davis_spin_coat}
{\sc \au{Reisfeld, B.}, \au{Bankoff, S.~G.} \& \au{Davis, S.~H.}} \yr{1991}
  \at{The dynamics and stability of thin liquid films during spin coating. {I}.
  {F}ilms with constant rates of evaporation or absorption}.  \jt{Journal of
  Applied Physics}  \bvol{70}~(10),  \pg{5258--5266}.

\bibitem[Renardy(1987)]{renardy_triple_layer}
{\sc \au{Renardy, Y.}} \yr{1987}  \at{Viscosity and density stratification in
  vertical {P}oiseuille flow}.  \jt{Physics of Fluids (1958-1988)}
  \bvol{30}~(6),  \pg{1638--1648}.

\bibitem[Saisorn \& Wongwises(2008)]{saisorn_exp_two_phase_map}
{\sc \au{Saisorn, S.} \& \au{Wongwises, S.}} \yr{2008}  \at{A review of
  two-phase gas-–liquid adiabatic flow characteristics in micro-channels}.
  \jt{Renewable and Sustainable Energy Reviews}  \bvol{12}~(3),  \pg{824 --
  838}.

\bibitem[Shlang {\em et~al.\/}(1985)Shlang, Sivashinsky, Babchin \&
  Frenkel]{shlang_double_nonlin}
{\sc \au{Shlang, T.}, \au{Sivashinsky, G.I.}, \au{Babchin, A.J.} \&
  \au{Frenkel, A.L.}} \yr{1985}  \at{Irregular wavy flow due to viscous
  stratification}.  \jt{J. Phys. France}  \bvol{46}~(6),  \pg{863--866}.

\bibitem[{Sivashinsky} \& {Michelson}(1980)]{sivashinsky}
{\sc \au{{Sivashinsky}, G.~I.} \& \au{{Michelson}, D.~M.}} \yr{1980}  \at{On
  irregular wavy flow of a liquid film down a vertical plane}.  \jt{Progress of
  Theoretical Physics}  \bvol{63},  \pg{2112--2114}.

\bibitem[Taitel \& Dukler(1976)]{taitel_anal_two_phase_map}
{\sc \au{Taitel, Y.} \& \au{Dukler, A.~E.}} \yr{1976}  \at{A model for
  predicting flow regime transitions in horizontal and near horizontal
  gas-liquid flow}.  \jt{AIChE Journal}  \bvol{22}~(1),  \pg{47--55}.

\bibitem[Takeshi(1999)]{ooshida_1999}
{\sc \au{Takeshi, O.}} \yr{1999}  \at{Surface equation of falling film flows
  with moderate {R}eynolds number and large but finite {W}eber number}.
  \jt{Physics of Fluids}  \bvol{11}~(11),  \pg{3247--3269}.

\bibitem[Talimi {\em et~al.\/}(2012)Talimi, Muzychka \&
  Kocabiyik]{talimi_numeric_two-phase}
{\sc \au{Talimi, V.}, \au{Muzychka, Y.S.} \& \au{Kocabiyik, S.}} \yr{2012}
  \at{A review on numerical studies of slug flow hydrodynamics and heat
  transfer in microtubes and microchannels}.  \jt{International Journal of
  Multiphase Flow}  \bvol{39},  \pg{88 -- 104}.

\bibitem[Triplett {\em et~al.\/}(1999)Triplett, Ghiaasiaan, Abdel-Khalik \&
  Sadowski]{triplett_anal_two_phase_map}
{\sc \au{Triplett, K.A.}, \au{Ghiaasiaan, S.M.}, \au{Abdel-Khalik, S.I.} \&
  \au{Sadowski, D.L.}} \yr{1999}  \at{Gas-–liquid two-phase flow in
  microchannels part {I}: two-phase flow patterns}.  \jt{International Journal
  of Multiphase Flow}  \bvol{25}~(3),  \pg{377 -- 394}.

\bibitem[VanHook {\em et~al.\/}(1997)VanHook, Schatz, Swift, McCormick \&
  Swinney]{vanhook_lw_exp_theo}
{\sc \au{VanHook, S.~J.}, \au{Schatz, M.~F.}, \au{Swift, J.~B.}, \au{McCormick,
  W.~D.} \& \au{Swinney, H.~L.}} \yr{1997}  \at{Long-wavelength
  surface-tension-driven {B}\'enard convection: experiment and theory}.
  \jt{Journal of Fluid Mechanics}  \bvol{345},  \pg{45--78}.

\bibitem[V\'ecsei {\em et~al.\/}(2014)V\'ecsei, Dietzel \& Hardt]{coupled_sos}
{\sc \au{V\'ecsei, M.}, \au{Dietzel, M.} \& \au{Hardt, S.}} \yr{2014}
  \at{Coupled self-organization: Thermal interaction between two liquid films
  undergoing long-wavelength instabilities}.  \jt{Phys. Rev. E}  \bvol{89},
  \pg{053018}.

\bibitem[Yih(1967)]{yih_viscous_strati}
{\sc \au{Yih, C.~S.}} \yr{1967}  \at{Instability due to viscosity
  stratification}.  \jt{Journal of Fluid Mechanics}  \bvol{27},  \pg{337--352}.

\end{thebibliography}
\end{document}